\documentclass[acmsmall]{acmart}
\usepackage{algorithmic}
\usepackage{algorithm}
\usepackage{array}
\usepackage{textcomp}
\usepackage{stfloats}
\usepackage{url}
\usepackage{verbatim}
\usepackage{graphicx}

\usepackage{amssymb}
\usepackage{amsmath,amsfonts}
\usepackage[mathscr]{eucal}
\usepackage{subfigure}
\usepackage{etoolbox}
\usepackage{multirow}
\usepackage{tabularx}
\usepackage{xcolor}
\usepackage{color}
\usepackage{booktabs}

\AtBeginDocument{%
  }

\setcopyright{acmcopyright}
\copyrightyear{2023}
\acmYear{2023}
\acmDOI{XXXXXXX.XXXXXXX}

\acmJournal{JACM}
\acmVolume{37}
\acmNumber{4}
\acmArticle{111}
\acmMonth{10}

\begin{document}

\title{Causal Inference with Complex Treatments: A Survey}



\author{Yingrong Wang}
\affiliation{%
  \institution{Zhejiang University}
  \city{Hangzhou}
  \country{China}}
\email{wangyingrong@zju.edu.cn}

\author{Haoxuan Li}
\affiliation{%
  \institution{Peking University}
  \city{Beijing}
  \country{China}}
\email{hxli@stu.pku.edu.cn}

\author{Minqin Zhu}
\affiliation{%
  \institution{Zhejiang University}
  \city{Hangzhou}
  \country{China}}
\email{minqinzhu@zju.edu.cn}

\author{Anpeng Wu}
\affiliation{%
  \institution{Zhejiang University}
  \city{Hangzhou}
  \country{China}}
\email{anpwu@zju.edu.cn}

\author{Ruoxuan Xiong}
\affiliation{%
  \institution{Emory University}
  \city{Atlanta}
  \country{US}}
\email{ruoxuan.xiong@emory.edu}

\author{Fei Wu}
\affiliation{%
  \institution{Zhejiang University}
  \city{Hangzhou}
  \country{China}}
\email{wufei@cs.zju.edu.cn}

\author{Kun Kuang}
\affiliation{%
  \institution{Zhejiang University}
  \city{Hangzhou}
  \country{China}}
\email{kunkuang@zju.edu.cn}


\renewcommand{\shortauthors}{Yingrong Wang et al.}

\begin{abstract}
  Causal inference plays an important role in explanatory analysis and decision making across various fields like statistics, marketing, health care, and education. Its main task is to estimate treatment effects and make intervention policies. Traditionally, most of the previous works typically focus on the binary treatment setting that there is only one treatment for a unit to adopt or not. However, in practice, the treatment can be much more complex, encompassing multi-valued, continuous, or bundle options. In this paper, we refer to these as complex treatments and systematically and comprehensively review the causal inference methods for addressing them. First, we formally revisit the problem definition, the basic assumptions, and their possible variations under specific conditions. Second, we sequentially review the related methods for multi-valued, continuous, and bundled treatment settings. In each situation, we tentatively divide the methods into two categories: those conforming to the unconfoundedness assumption and those violating it. Subsequently, we discuss the available datasets and open-source codes. Finally, we provide a brief summary of these works and suggest potential directions for future research.
\end{abstract}



\received{20 February 2007}
\received[revised]{12 March 2009}
\received[accepted]{5 June 2009}

\maketitle

\section{Introduction}\label{sec:intro}
Causal inference has extensive applications in many domains such as statistics~\cite{statistics}, marketing~\cite{marketing}, epidemiology~\cite{epidemiology}, education~\cite{education}, recommendation system~\cite{recommendation}, etc. Although association models have gained interest in these domains, they are limited to specific settings with independent and identically distributed (i.i.d.) data. In contrast, causal methods have already considered this data distribution gap when determining the actual impact of a treatment or intervention on a particular outcome. Formally, the treatment effect refers to the difference in outcome that would have been resulted if a treatment of interest had been taken, compared to the circumstance where it had not. Such estimation is helpful in not only effect measurement but also some downstream tasks like prediction, decision making, feature selection, and explanatory analysis.

The key challenge when estimating the treatment effect is to control the confounding bias. It means that confounders may simultaneously affect both the independent (treatment) and dependent (outcome) variables, thus leading to incorrect estimation of causalities and treatment effect. For instance, age is a confounder when studying the effect of smoking on lung cancer, since the age could both affect whether one smokes and the probability of having lung cancer. 

In practice, the gold standard methods for estimating the treatment effect are randomized controlled trials (RCTs), where the treatment is randomly assigned to units. However, it is usually expensive to conduct RCTs and it is difficult for RCTs to figure out the effects of complex treatments. Moreover, it may conflict with ethical principles, especially in medical scenarios. For example, when studying the effect on mortality of a certain drug, it is immoral and illegal to force patients to receive a treatment or not. Therefore, many recent work focuses on how to precisely estimate the treatment effect from observational data that are naturally collected. In this paper, we focus on surveying the methods of causal inference in observational studies.

To formally study this problem, we adopt the widely used potential outcome framework~\cite{rubin1974estimating} in causal inference literature~\cite{neyman1990application,fisher1935design,holland1986statistics}. A variety of methods
have emerged, including propensity-based methods, representation-based methods, generative modeling methods, etc. Propensity score~\cite{rosenbaum1983central} estimates the conditional probability of a sample adopting a particular treatment with given covariate. On the basis of propensity score, methods like matching~\cite{matching}, stratification~\cite{stratification}, and re-weighting~\cite{re-weighting} are proposed to control the confounding bias. With further consideration of the selection bias, the balancing property of propensity are leveraged to mimic the randomization in observational data. In order to control the distribution of variables from different units, balancing methods are developed, including entropy balancing~\cite{entropy}, covariate balancing propensity score~\cite{CBPS}, approximate residual balancing~\cite{approximate}, and kernel balancing~\cite{wong2018kernel}. With the progress of deep learning, recent studies apply neural networks to learn representation for the covairates of a unit, followed by a hypothesis network to infer the potential outcome. These methods encourage similarity between the representations of the two groups, which benefits the distribution balance. There are various examples, including Balancing Neural Network (BNN)~\cite{BLR}, CounterFactual Regression (CFR)~\cite{CFR}, CounterFactual Regression with Importance Sampling Weights (CFR-ISW)~\cite{CFR-ISW}, Dragonnet~\cite{Dragonnet}, etc. Besides, there are several methods that make use of multi-task learning~\cite{alaa2017bayesian,alaa2017deep} and meta learning~\cite{kunzel2019metalearners,nie2021quasi}. Generative modeling methods are another kind of mainstream approaches, which utilize generative adversarial network (GAN)~\cite{GAN} or variational auto-encoder (VAE)~\cite{VAE}. GANITE~\cite{GANITE} is a representative of the former and directly generates the potential outcomes via generators. As for the latter, the main idea is to obtain a latent variable or embedding for the target of interest by a reconstruction loss and distribution discrepancy measurement. Specifically, such target in Causal Effect VAE (CEVAE)~\cite{CEVAE} is the unmeasured confounders, which are recovered as a latent variable and used in the following estimation of causal effect.

\begin{figure}[t]
	\centering
	\includegraphics[width=0.9\textwidth]{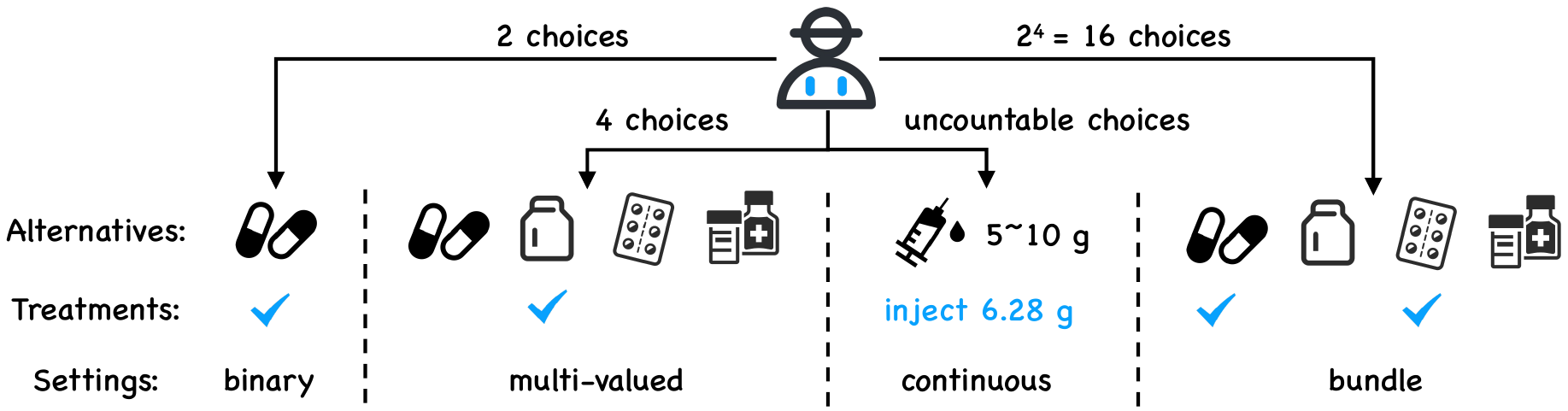}
	\vspace{-2mm}
	\caption{An example of complex treatments.}
	\label{fig:complex-treatment}
    \vspace{-3mm}
\end{figure}

Previous effects mainly focus on the setting of binary treatment, meaning that there is only a single treatment to be adopted or not. However, the treatment could be multi-valued, bundle, continuous, or even more complex in piratical applications. We give an example under the circumstance of making decision on the medications, which is illustrated in Fig.~\ref{fig:complex-treatment}. If the treatment is a single variable, the patient could decide whether to take a certain drug or not (binary), choose one from multiple alternatives (multi-valued), or even consider the injection dosage (continuous). On the other hand, the treatment can also consist of multiple variables, and the patient is to consider a combination of several drugs (bundle).

Therefore, causal inference with complex treatments has drawn increasing attention in recent years. Generalized propensity score (GPS)~\cite{GPS} is an extension of the propensity score. Many methods based on GPS are proposed to estimate the causal effect of multi-valued treatment~\cite{multiGPS} and continuous treatment~\cite{continuousGPS}. Similarly, the representation-based methods also play an important role under the settings of complex treatments. For instance, CFR~\cite{CFR} is extended to the situations of multi-valued treatment as MEMENTO~\cite{MEMENTO} and bundle treatment as Regret Minimization Network (RMNet)~\cite{Regret}, respectively. Dose Response Network (DRNet)~\cite{DRNet} and Varying Coefficient Network (VCNet)~\cite{VCNet} are also good examples of the representation-based methods for continuous treatment. As for the generative modeling methods using GAN, GANITE can be naturally applied for estimating the effect of treatment with multiple values, as long as changing the task of discriminators as multi-class classification. SCIGAN~\cite{SciGAN} is a further exploration under the circumstance of continuous setting. Researchers have made good use of VAE as well, developing Task Embedding based Causal Effect VAE (TECE-VAE)~\cite{TECE-VAE} and Variational Sample Re-weighting (VSR)~\cite{VSR} for bundle treatment, together with Identifiable treatment-conditional VAE (Intact-VAE)~\cite{Intact-VAE} for continuous treatment. Apart from the three mainstream approaches, MetaITE~\cite{meta} provides another solution for treatment with multiple values. Specifically, it regards those treatment groups with sufficient samples as source domains and thus trains a meta-learner. On the other hand, the group with limited samples is treated as a target domain for model update.

\begin{figure}[b]
    \vspace{-5mm}
    \centering
    \subfigure[Observed]{
        \includegraphics[scale=0.43]{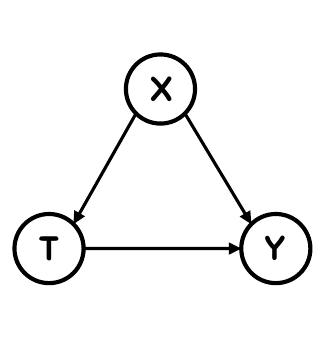}
        \vspace{-3mm}
        \label{fig:unconfoundedness}
    }
    \hspace{5mm}
    \subfigure[Unobserved]{
        \includegraphics[scale=0.43]{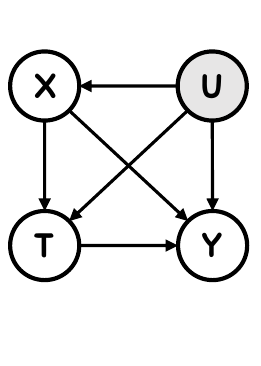}
        \vspace{-3mm}
        \label{fig:unobserved}
    }
    \hspace{5mm}
    \subfigure[Proxy Variable]{
        \includegraphics[scale=0.43]{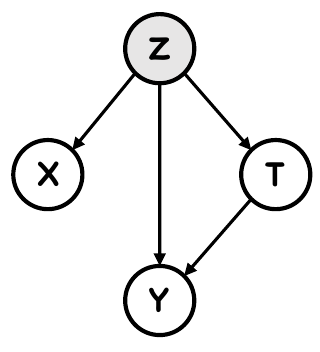}
        \vspace{-3mm}
        \label{fig:proxy}
    }
    \hspace{5mm}
    \subfigure[Instrumental Variable]{
        \includegraphics[scale=0.43]{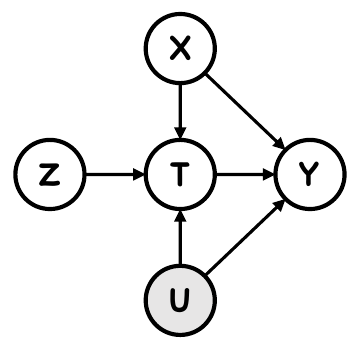}
        \vspace{-3mm}
        \label{fig:IV}
    }
    \vspace{-3mm}
    \caption{Potential outcome frameworks w/o unobserved confounders (marked in shadow).}
    \label{fig:framework}
\end{figure}

Due to limitations of information collection, there may exist unobserved confounder. As shown in Fig.~\ref{fig:unobserved}, the unobserved confounder $U$ in shadow means that it can not be measured or does not appear in the dataset. It is also called confounder because it simultaneously affects treatment $T$ and outcome $Y$. Note that $U$ may have a causal link with confounder $X$ that can be observed. One method to address this issue is to find a proxy variable as a subsititue of the unmeasured confounder. As shown in Fig.~\ref{fig:proxy}, $Z$ is recovered to serve as the proxy of the unmeasured condounders, which affects $X$ and $T$ at the same time. A requirement is that $X$ is independent to $T$ given $Z$. Many works make effort to find such a proxy variable, like Multiple Causal Estimation via Information (MCEI)~\cite{multiProxy} for multi-valued treatment, deconfounders~\cite{deconfounder-2018,deconfounder-2021} for bundle treatment, and Deep Feature Proxy Variable (DFPV)~\cite{DFPV} for continuous treatment. Additionally, instrumental variable (IV) is also widely used in this situation, which is illustrated in Fig.~\ref{fig:IV}. Given $X$, the instrumental variable $Z$ is beneficial to the identification of $T\rightarrow Y$. DeepIV~\cite{DeepIV} and IV using Producing Kernel Hilbert Spaces (RKHS)~\cite{singh2020kernel} are two instances for such instrumental variable methods.

\begin{figure}[t]
	\centering
	\includegraphics[width=1\textwidth]{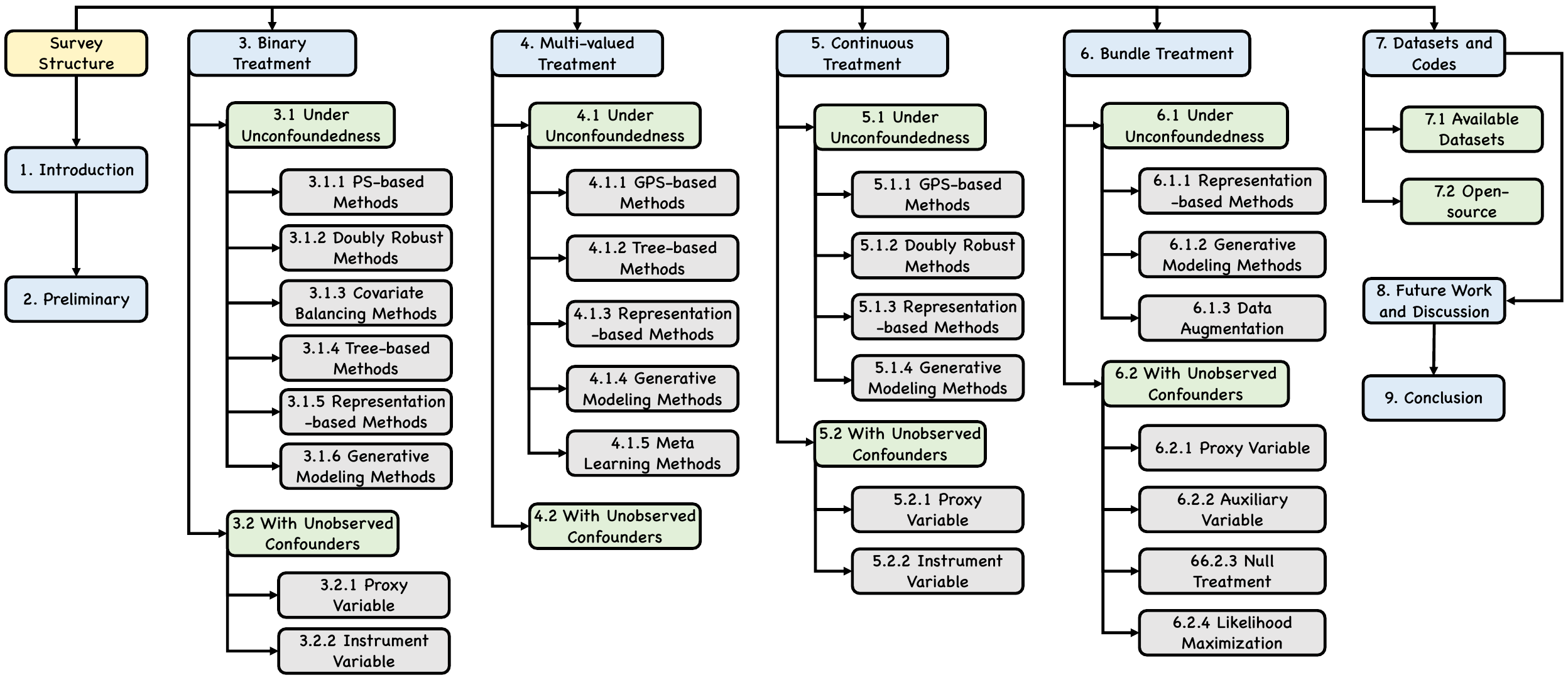}
	\vspace{-8mm}
	\caption{Outline of the survey.}
	\label{fig:outline}
    \vspace{-5mm}
\end{figure}

There are several surveys in the causal inference community, such as the two focused on binary treatment~\cite{causal-survey,guo2021causality}, the work that concludes the instrumental variable methods~\cite{iv-survey}, and the one discussing the matching methods for multi-valued treatment~\cite{multiGPS}. However, the problem of estimating causal effect of complex treatments is rarely discussed, which is common and important in practical applications. In this paper, we provide a comprehensive review on methods with complex treatments under the potential outcome framework. We clarify the problem setups of the multi-valued, continuous, and bundle treatment settings, and distinguish the similarities and differences among them. We give a brief introduction of some representative methods, together with common experimental datasets and details. Key challenges induced by the distinct treatment settings will be further discussed as well.

\textbf{Paper organization.} Architecture of this paper is illustrated in Fig.~\ref{fig:outline}. Preliminaries of causal inference with complex treatments will be introduced in Section~\ref{sec:preliminary}. Methods catered for the setting of binary treatment are listed in Section~\ref{sec:binary}, multi-valued treatment in Section~\ref{sec:multi}, continuous treatment in Section~\ref{sec:continuous}, and bundle treatment in Section~\ref{sec:bundle}. Afterwards, we collect several available datasets and open-source codes in Section~\ref{sec:experiment}. We give a further discussion about the directions of future works in Section~\ref{sec:discussion}, and a brief conclusion in Section~\ref{sec:conclusion}.

\begin{table}[t]
\small
\centering
\caption{Important notations.}
\vspace{-2mm}
\label{tab:notaton}
\begin{tabular}{cll}
\toprule
\textbf{Notation} & \textbf{Definition or Domain} & \textbf{Explanation} \\
\midrule
$n$ & $\in\mathbb{R}_+$ & number of units \\
$i$ & $\in\{0,\dots,n-1\}$ & indicator of each unit \\
$d$ & $\in\mathbb{R}_+$ & dimension of covariate \\
$\mathcal{X}$ & $\subset\mathbb{R}^d$ & support of covariate \\
$X_i$ & $\in\mathcal{X}$ & covariate of unit $i$ \\
$\mathcal{T}^{bin}$ & $=\{0,1\}$ & support of binary treatment \\
$m$ & $\in\mathbb{R}_+$ & number of multiple treatments \\
$\mathcal{T}^{mul}$ & $=\{0,1,\dots,m\}$ & support of multi-valued treatment \\
$\mathcal{T}^{bun}$ & $\subset\{0,1\}^m$ & support of bundle treatment \\
$T_i$ & $\in\mathcal{T}$ & treatment of unit $i$ \\
$\mathcal{Y}$ & $\subset\mathbb{R}$ & support of outcome \\
$Y_i$ & $\in\mathcal{Y}$ & factual outcome of unit $i$ \\
$Y_i(t)$ & $=Y_i(T_i=t)$ & potential outcome of unit $i$ if receiving treatment $t$ \\
$ITE_i$ & $=Y_i(t)-Y_i(0)$ & individual treatment effect of unit $i$ \\
$ATE$ & $=\mathbb{E}[Y(t)-Y(0)]$ & average treatment effect of all units \\
$CATE$ & $=\mathbb{E}[Y(t)-Y(0)|X=x]$ & conditional average treatment effect \\
$IDRF_i$ & $=Y_i(t)$ & individual dose-response function of unit $i$ \\
$ADRF$ & $=\mathbb{E}[Y(t)]$ & average dose-response function of all units \\
$HDRF$ & $=\mathbb{E}[Y(t)|X=x]$ & heterogeneous dose-response function \\
$U$ & & unmeasured confounders \\
$Z$ & & instrumental variable or proxy \\
$r(X)$ & $=P(T=1|X)$ & propensity score of binary treatment \\
$r(T,X)$ & $=f_{T|X}(T|X)$ & generalized propensity score of multiple treatments \\
\bottomrule
\vspace{-7mm}
\end{tabular}
\end{table}

\section{Preliminary}~\label{sec:preliminary}
We first introduce the basic setups in the case of binary treatment. Suppose there is a random sampling of $n$ units from a population $P$. We denote the covariate of each unit $i$ as $X_i\in\mathcal{X}\subset\mathbb{R}^d$, and the assigned treatment as $T_i\in\mathcal{T}^{bin}=\{0,1\}$. When there exist $m$ discrete treatments, we rewrite the treatment as $T_i\in\mathcal{T}^{mul}=\{0,1,\dots,m\}$ for the multi-valued setting, and $T_i\in\mathcal{T}^{bun}\subset\{0,1\}^m$ for the bundle setting. As for the continuous treatment, it can be denoted as $T_i\in\mathcal{T}^{con}\subset\mathbb{R}$. The outcome of unit $i$ receiving a specific treatment $T_i$ is $Y_i\in\mathcal{Y}\subset\mathbb{R}$. Note that we consider the circumstance of continuous outcome in this paper. We adopt the potential outcome framework~\cite{rubin1974estimating,neyman1990application} in causal inference. For generality, let $Y_i(t)$ and $Y_i(0)$ be the outcome of receiving treatment $T_i=t$ and no treatment $T_i=0$. Only one of them can be observed in the dataset while the other is obtained by counterfactual prediction, which is known as the fundamental problem of causal inference~\cite{holland1986statistics,morgan2015counterfactuals}. \textcolor{black}{In Table~\ref{tab:notaton}, we conclude some important notations that are commonly used in this community.}

Three basic assumptions are proposed to make sure the identifiablity of treatment effect estimation. The first one is \textit{stable unit treatment value assumption (SUTVA)}, which contains considerations from two aspects. On the one hand, units are independent with each other, having no influence on others' outcomes. On the other hand, there are no alternative forms of a treatment, meaning that the observed outcome is the potential outcome corresponding to the assigned treatment. The second one is \textit{unconfoundedness assumption}, also called \textit{ignorability}, that $Y_i(T_i=t)\perp\!\!\!\perp T_i\,|\,X_i$. This assumption guarantees that there are no unobserved confounders. The third one is called \textit{positivity} or \textit{overlap assumption}, i.e. $P(T_i=t|X_i=x)>0, \forall\,t, x$. It is proposed in case of the condition that there is no unit applying a certain treatment. Note that the \textit{unconfoundedness} and \textit{positivity} assumption are collectively referred to as \textit{strong ignorability assumption}~\cite{imbens2015causal}.

On the basis of the counterfactual outcome and the assumptions above, the individual treatment effect (ITE) of unit $i$ can be measured as $ITE_i = Y_i(T_i=t)-Y_i(T_i=0)$. If $ITE_i > 0$, then the treatment $t$ is more beneficial than receiving no treatment, and vice versa.  As for the whole population, we use average treatment effect (ATE) to quantify the treatment effect, i.e. $ATE = \mathbb{E}[Y(T=t)-Y(T=0)]$. In order to study the treatment effect on samples with particular characteristics, we can pick out a subgroup and the treatment effect on them is called conditional average treatment effect (CATE), i.e. $CATE = \mathbb{E}[Y(T=t)-Y(T=0)|X=x]$. This measurement also plays an important role when treatment effect varies significantly across different subgroups, which is also known as the heterogeneous treatment effect (HTE).

When it comes to the continuous treatment, a \textit{smoothness assumption} is proposed that the potential outcome $Y(T=t)$ is a smooth response to treatment $T=t$. Besides, the \textit{unconfoundedness assumption} for continuous treatment should be rewritten as $\{Y(t)\} \perp \!\!\! \perp T\,|\,X, \forall t\in T$. There is also a weaker version that $Y(t) \perp \!\!\! \perp T\,|\,X, \forall t\in T$. The key for measuring effects of continuous treatment is the dose-response function, and there are various measurements~\cite{galvao2015uniformly,farrell2020deep}. Similar to ITE, the formal definition of individual dose-response function (IDRF) is given as $IDRF_i=Y_i(T_i=t)$. The average dose-response function (ADRF) quantifies the causal effect on the whole population by $ADRF=\mathbb{E}[Y(T=t)]$. For the heterogeneous treatment effect under the continuous treatment setting, heterogeneous dose-response function (HDRF) is proposed as $HDRF=\mathbb{E}[Y(T=t|X=x)]$.

\section{Binary Treatment}~\label{sec:binary}
Considering that many methods tackling with complex treatments are developed from models of the binary setting, so we will give a brief introduction of those binary treatment methodologies at first. This section is to serve as the support of background knowledge for the following parts.

\subsection{Under Unconfoundedness}
Firstly, we will introduce approaches when the three basic assumptions hold.

\subsubsection{Propensity score (PS)-based Methods}
Propensity score is one of the most common methods used for settings of binary treatment. Its definition can be described as the following equation, which refers to the conditional probability of a unit receiving treatment $T$ when given covariates $X$:
\begin{equation}\label{eq:ps}
r(X)=P(T=1|X).
\end{equation}
In this way, we are able to figure out how the covariates $X$ affect the assignment of $T$. Based on this, many approaches are proposed to alleviate the problem of confounding bias to a certain extent, such as matching~\cite{matching}, stratification~\cite{stratification}, and re-weighting~\cite{re-weighting}. The main idea of \textbf{matching} methods is to design a distance matrix and matching algorithm using propensity score, so as to determine the matched pairs across the treated and control groups. As for a unit $i$ whose matched neighbours from the opposite group are denoted as $\mathcal{J}(i)$, we can estimate the counterfactual outcomes from the observed outcomes by:
\begin{equation}
    \hat{Y}_i(1-t)=\frac{1}{|\mathcal{J}(i)|}\sum_{j\in\mathcal{J}(i)}Y_j(t).
\end{equation}
Various designs about the distance measurement and matching algorithm are discussed in the survey~\cite{matching-survey}. \textbf{Stratification} methods, also named \textit{sub-classification} or \textit{blocking}, are aimed to split the entire group into several homogeneous subgroups (blocks), within each those units in the treated group and the control group are similar. 
How to split all the samples is based on the propensity score as well. Suppose there are $J$ blocks, we can estimate the ATE as follows:
\begin{equation}
    ATE_{strat}=\sum_{j=1}^J\frac{n_j}{n}\left[\bar{Y}_t(j)-\bar{Y}_c(j)\right],
\end{equation}
where $n_j$ is the number of samples in the $j$-th block, and $\bar{Y}_t(j)$ and $\bar{Y}_c(j)$ are the average outcomes of the treated group and control group, respectively. When it comes to \textbf{re-weighting} methods, they are focused on assigning appropriate weight to each unit in order to construct a new population where distributions of the treated group and control group are similar. Take Inverse propensity weighting (IPW)~\cite{re-weighting,rosenbaum1983central} for example, the sample weights $w$ can be defined as:
\begin{equation}
    w=\frac{T}{r(X)}+\frac{1-T}{1-r(X)}.
\end{equation}
Therefore, ATE can be estimated with observed outcomes $Y_i$ through:
\begin{equation}\label{eq:ipw}
    ATE_{IPW}=\frac{1}{n}\sum_{i=1}^n\frac{T_i Y_i}{r(x_i)}-\frac{1}{n}\sum_{i=1}^n\frac{(1-T_i)Y_i}{1-r(x_i)}.
\end{equation}

\subsubsection{Doubly Robust Methods}\label{sec:dr}
Although the PS-based methods have been widely developed, one key concern is that all these methods heavily rely on the correctness of estimating the propensity score. Take IPW for instance, even the slight misspecification of propensity scores can cause significant error when estimating $ATE_{IPW}$. To address this issue, \textbf{Doubly Robust (DR)}~\cite{DR1}, also called Augmented IPW, combines the regression of propensity score and potential outcome:
\begin{equation}
    ATE_{DR}=\frac{1}{n}\sum_{i=1}^n\left[\hat{m}(1,x_i)-\hat{m}(0,x_i)+\frac{T_i(Y_i-\hat{m}(1,x_i))}{r(x_i)}-\frac{(1-T_i)(Y_i-\hat{m}(0,x_i))}{1-r(x_i)}\right],
\end{equation}
where $\hat{m}(\cdot)$ is the regression model that estimates the potential outcomes. In this way, even one of $\hat{m}(\cdot)$ or $r(\cdot)$ has poor performance, the overall the estimator is still robust. Although the doubly robust method is initially proposed as an improvement of IPW, it later evolves into a robust framework between double regression models, inspiring many other new works.

\subsubsection{Covariate Balancing Methods}

Considering that the regression of propensity scores often rely on model specification, researchers propose alternative approaches that directly adjust the covariate distribution of two groups so as to control the selection bias. It is like simulating the randomization process with observational data for the purpose of achieving $T_i\ \!\!\!\perp X_i$. The main idea is to assign weights to each sample to ensure the reweighted groups satisfy the balance constraints, that is aligning the first-order moment of sample covariates between the treated group and the control group. \textbf{Entropy balancing}~\cite{entropy} method determines the reweighting scheme by minimizing the entropy divergence between distributions of the two groups. As for \textbf{Covariate balancing propensity score (CBPS)}~\cite{CBPS}, it utilizes the balancing property of propensity score, i.e. $T_i\ \!\!\!\perp X_i | r(X_i)$, to improve its estimation. To be specific, the propensity scores are solved by:
\begin{equation}
    \mathbb{E}\left[\frac{T_i\tilde{X}_i}{r(X_i)}-\frac{(1-T_i)\tilde{X}_i}{1-r(X_i)}\right]=0.
\end{equation}
where $w_i$ represents the weights of $X_i$, and $\tilde{X}_i=w_iX_i$ is the adjusted covariates after reweighting. \textbf{Approximate residual balancing}~\cite{approximate} is another method that combines the idea of doubly robust. Specifically, it combines balancing weights learning, propensity score regression and potential outcome estimation together. \textbf{Kernel balancing}~\cite{wong2018kernel} is proposed in recent years, which attains uniform approximate balance for covariate functions in a reproducing-kernel Hilbert space.

\subsubsection{Tree-based Methods}

The tree structure, such as the Classification And Regression Tree (CART)~\cite{breiman2017classification}, is also widely utilized to estimate heterogeneity in causal effects. To be specific, the process of partitioning the whole population into several sub-groups can be simulated through the tree splitting. Afterwards, the treatment effects can be estimated according to other samples that fall into the same leaf node of the query sample. \textbf{Bayesian Additive Regression Trees (BART)}~\cite{BART} is a Bayesian ensemble method that models the mean outcome given predictors by a sum of trees:
\begin{equation}
f(t,X_i)=\mathbb{E}(Y_i|T_i=t,X_i)=\Phi\left\{ \sum_{j=1}^M g_j(t,X_i;R_j,\theta_j) \right\},
\end{equation}
where $\Phi$ refers to the standard normal cumulative distribution function, $R_j$ denotes the $j$-th regression tree, and $\theta_j$ is a set of parameter values associated with the terminal nodes of it. $g_j(t,X_i)$ represents the mean assigned to the node in the $j$-th regression tree associated with covariate $X_i$ and treatment $t$. The causal estimand of interest can be estimated by contrasting the imputed potential outcomes between treatment groups. \textcolor{black}{It can be seen that the tree methods is naturally applicable to tackle with the causal effect estimation of multi-valued treatment.}

\subsubsection{Representation-based Methods}

\textbf{BNN}~\cite{BLR} and \textbf{CFR}~\cite{CFR} are two of the most classic methods based on invariant representation. Generally speaking, this kind of method includes a representation network $\Phi(x)$ to learn the universal representation for samples from both groups, together with a hypothesis network $h(\Phi)$ to predict potential outcomes. Considering that the populations of different treatment groups are supposed to be similar or balanced, restrictions to minimize their discrepancy are applied as well. Therefore, the basic objective function of representation-based methods can be concluded as:
\begin{equation}
    \mathcal{L}=\mathcal{L}(h)+\mathcal{L}(\Phi)+\mathcal{R}.
\end{equation}
The first term refers to the prediction error of hypothesis network(s). The second term is a quantitative measurement for the discrepancy distance between the distributions of the treated group and the control group. The last one is an optional regularization term about model complexity. Take CFR for example, its objective function is given in the following equation:
\begin{equation}\label{eq:CRF}
\min\limits_{\stackrel{h,\Phi}{\lVert\Phi\rVert}}\frac{1}{n}\sum_{i=1}^n w_i\cdot L\left(h(\Phi(X_i),T_i),Y_i\right)+\alpha\cdot{\rm IPM_G}\left(\{\Phi(X_i)\}_{i:T_i=0},\{\Phi(X_i)\}_{i:T_i=1}\right)+\lambda\cdot\mathcal{R}(h),
\end{equation}
where $w_i=\frac{T_i}{2u}+\frac{1-T_i}{2(1-u)}$ and $u=\frac{1}{n}\sum_{i=1}^nT_i$.

\subsubsection{Generative Modeling Methods} \textbf{GANITE~\cite{GANITE}} introduces the idea of Generative Adversarial Network (GAN)~\cite{GAN} into the causal inference community. It is composed of a counterfactual block and an ITE block, and in each block there is a separate GAN structure. In the counterfactual block, the generator is to fill up all the missing counterfactual outcomes $Y_i^{CF}$ while the discriminator is to decide whether the potential outcome is the real data $Y_i^F$ or the fake ones derived from the generator. In this way, a "complete" dataset can be obtained. As for the ITE block, there is a generator to estimate the outcome $\widehat{ITE_i}$ given the covariate $X_i$, and a discriminator aimed at judging whether its input is $ITE_i$ form the dataset after imputation or $\widehat{ITE_i}$ from the generator. Ultimately, the two generators can give accurate estimations of $Y_i^{CF}$ and $\widehat{ITE_i}$. \textcolor{black}{Despite of the considerable performance achieved, methods based on GAN still lack theoretical guarantees.}

\subsection{With Unobserved Confounders}
Proxy variable and instrumental variable are strong tools when there exist unobserved confounders. \textcolor{black}{Although these methods are initially proposed to tackle with binary treatment, they can be naturally developed to solve the problem of continuous treatment. Therefore, we just give a brief introduction of these concepts in this section, and discuss some concrete methods in Section~\ref{sec:continuous-unobserve}.}

\subsubsection{Proxy Variable}
It has been studied for a long time as bias analysis~\cite{bias-study1,bias-study2}. The main idea of proxy is briefly described in Section~\ref{sec:intro} and Fig.~\ref{fig:proxy}. We tentatively divide the proxy variable methods into two categories, including negative controls and generative modeling methods.

\textbf{Negative Controls.} A recent study~\cite{miao2018identifying} summarizes the previous works on proxy and proposes the concept called negative control variables, which can be divided into negative control outcome (NCO) and negative control exposure (NCE). Generally speaking, a variable is NCO (denoted as $O$) if $O\perp\!\!\!\perp T|(U,X)$ and $O\not\perp\!\!\!\perp (U,X)$, or it is NCE (denoted as $E$) if $E\perp\!\!\!\perp Y|(U,X,T)$ and $E\perp\!\!\!\perp O|(U,X,T)$. In this case, the basic assumptions should be updated and there are additional assumptions.
\begin{itemize}
    \item \textit{SUTVA} for proxy. If treatment $T=t$ and NCE $E=e$, then the outcome and NCO is unique, i.e. $Y=Y(t,e) {\rm \ and\ } O=O(t,e)$.
    \item \textit{Positivity} for proxy. If $f(U,X)>0$, then we have $f(T,E|U,X)\in(0,1)$. Thereinto, $f(U,X)$ is the joint distribution of confounders $U$ and $X$, and $f(T,E|U,X)$ is the joint conditional density of treatment $T$ and NCE $E$ given confounders $U$ and $X$.
    \item \textit{Confounding bridge.} There exists at least one function $b(O,T)$ for all $T$ that could satisfy the condition $\mathbb{E}[Y|U,T]=\mathbb{E}[b(O,T)|U,T]$.
    \item \textit{Latent ignorability.} $Y(t)\perp\!\!\!\perp T|U, \forall t\in T$.
\end{itemize}

The latent ignorability assumption~\cite{miao2018confounding} can be viewed as the generalization of the unconfoundedness assumption. With the help of two additional assumptions mentioned above, we can identify the average causal effect as $\mathbb{E}[Y(t)]=\mathbb{E}[b(O,t)],\forall t\in T$. Afterwards, the confounding bridge function can be identified by NCE by $\mathbb{E}[Y|E,T]=\mathbb{E}[b(O,T)|E,T]$. Therefore, the definition of ADRF can be rewritten as $ADRF=\int b(T,O)d(T,O)$.

\textbf{Generative Modeling Methods.} Variational Auto-Encoder (VAE)~\cite{VAE} is a popular method for learning representation of the unobserved confounders.
\textbf{CEVAE}~\cite{CEVAE} is a classic method that leverages VAE to learn representations of $Z$ in causal inference for binary treatment. It consists of an inference network and a model network that are all derived from TARNet~\cite{CFR}, whose objective is to deduce the nonlinear relationship between $X$ and $(Z,y,t)$ so as to obtain the approximate solution of $P(Z,X,t,y)$. The variational lower bound is:
\begin{equation}
\mathcal{L}=\sum_{i=1}^n\mathbb{E}_{q(z_i|x_i,t_i.y_i)}[\log p(x_i,t_i|z_i)\ +\log p(y_i|t_i,z_i)+\log p(z_i)-\log q(z_i|x_i,t_i,y_i)],
\end{equation}
where the first two terms refer to the reconstruction loss and the last two represent the KL divergence. For each sample $X$, it first goes through the inference network to obtain $P(Z|X,y,t)$. Putting it into the model network gives the value of $P(y|t=1,X)$ and $P(y|t=0,X)$, respectively. ITE is the difference between them. Note that the intention of VAE used here is not to generate samples but to offer better representations of the hidden confounders for the causal estimand.

\subsubsection{Instrumental Variable}
It is a powerful tool for causal inference with unobserved confounders. A variable $Z$ is regarded as an IV if all the following conditions could be satisfied:
\begin{enumerate}
    \item \textbf{Relevance.} $Z$ is related to $X$, i.e., $Z\not\perp\!\!\!\perp X$.
    \item \textbf{Exclusion.} $Z$ affects $Y$ only through $T$, i.e. $Z\perp\!\!\!\perp Y|(T,U)$.
    \item \textbf{Unconfounded instrument.} $Z$ is independent of the unobserved confounders $U$, i.e. $Z\perp\!\!\!\perp U$.
\end{enumerate}
Generally speaking, it is hard to determine suitable IVs for the treatments of interest in reality and often requires professional knowledge. Therefore, an IV that meets all the conditions above is called valid IV, while weak IV refers to those have weak correlations with the treatment. Even worse, an IV is regarded as invalid IV with which the aforementioned assumptions will be violated.

\subsection{Conclusion and Discussion}
Methods of binary treatment are divided into two main types, depending on the existence of unobserved confounders. Under the \textit{unconfoundedness} assumption, researchers focus on eliminating the effects of $X$ on $T$. Propensity score is such an intuitive approach. However, model misspecification is a key concern, and thus lead to the development of covariate balancing methods and representation-based methods. When considering the existence of unobserved confounders, approaches using proxy variable and instrumental variable are tow mainstreams. Proxy variables are often recovered from observed data (such as treatment assignment), and then serve as the substitute of unmeasured confounders for counterfactual prediction. However, proxies are supposed to be sufficient to cover all the unmeasured confounders, making the recovery of proxy variable to be a challenging issue. Instrumental variables are auxiliary information to help estimate the influence of $T$ on $Y$, but how to provide valid IVs is also a hard problem that often needs expert knowledge.

\begin{figure}[t]
\vspace{-3mm}
	\centering
		\includegraphics[width=0.7\textwidth]{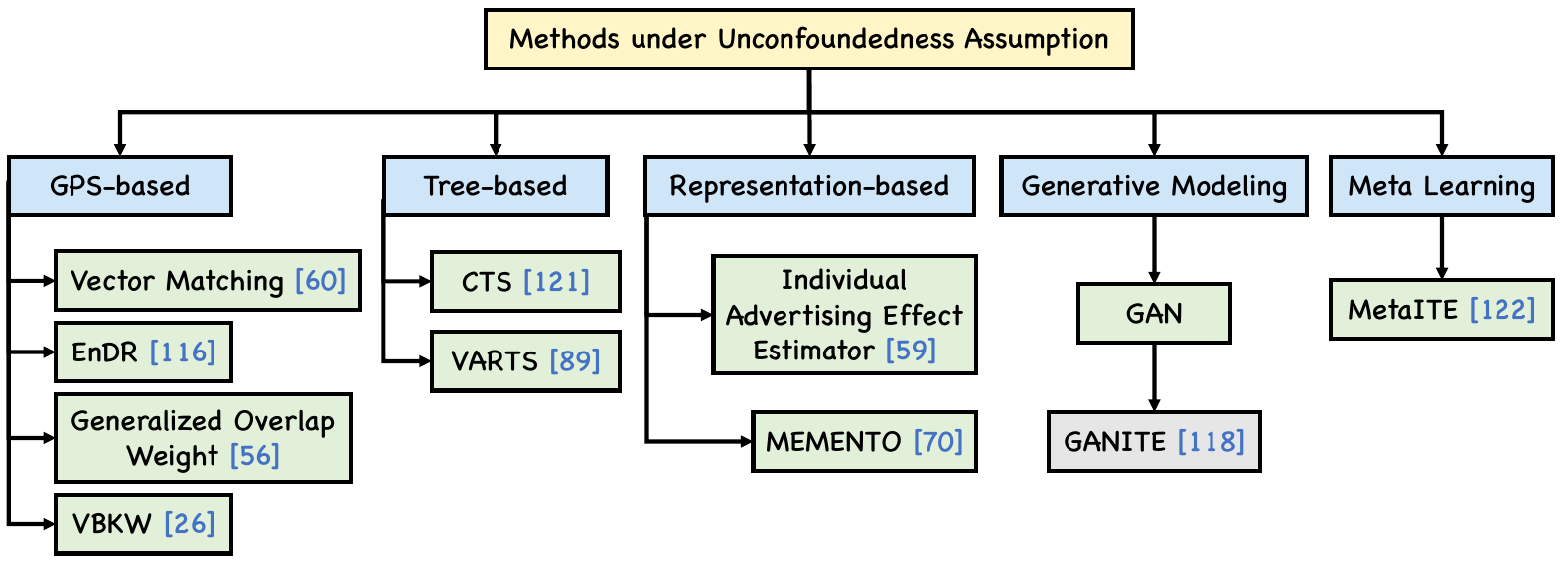}
	\vspace{-3mm}
	\caption{Categorization of multi-valued treatment methods without unobserved confounders.}
	\label{fig:multi}
\end{figure}

\section{Multi-valued Treatment}~\label{sec:multi}
In this section, we introduce relevant methods that estimate the causal effect of multi-valued treatment in two cases. The first case is that the \textit{unconfoundedness} assumption holds, and the second case is that there exist unobserved confounders.

\subsection{Under Unconfoundedness}\label{sec:multi-basic}
Existing works for multi-valued treatment under the \textit{unconfoundedness assumption} are organized in Fig.~\ref{fig:multi}. They can be further divided into $5$ categories: GPS-based methods, tree-based models, representation-based methods, generative modeling methods, and meta learning methods.

\subsubsection{GPS-based Methods}\label{sec:GPS}
Generalized propensity score (GPS)~\cite{GPS} is a extension derived from PS, whose definition is given in Eq. (\ref{eq:ps}). GPS is proposed as a solution for the settings of multiple treatments with discrete values and its expression is given as below:
\begin{equation}\label{eq:gps}
r(T,X)=f_{T|X}(T|X),
\end{equation}
where $f_{T|X}(T|X)$ means the conditional density of $T$ given $X$. Suppose there are $m$ treatments, then the GPS can be rewritten in the form of a vector as $R(X)=\left(r(t_1,X),\dots,r(t_m,X))\right)$. \textcolor{black}{Afterwards, many approaches using PS in the binary treatment setting, such as matching and doubly robust, are also extended to the multi-valued treatment setting by using GPS.}

\textbf{Vector Matching~\cite{multiGPS}} is proposed to match subjects with similar $R(X)$. Specifically, a multinomial regression model is applied to determine a common support region for multiple treatments. For each treatment $t \in \mathcal{T}$, we can get two bounds:
\begin{equation}
r(t,X)^{(low)}=\max\left(\min\left(r(t,X|T=t_1)\right),\dots,\min\left(r(t,X|T=t_m)\right) \right),
\end{equation}
\begin{equation}
r(t,X)^{(high)}=\min\left(\max\left(r(t,X|T=t_1)\right),\dots,\max\left(r(t,X|T=t_m)\right) \right),
\end{equation}
where $r(t,X|T=l)$ refers to the treatment assignment probability for $t$ among those subjects that received treatment $l$. Subjects with $r(t,X) \notin (r(t,X)^{(low)}, r(t,X)^{(high)})\ \forall t \in \mathcal{T}$ will be discarded, followed by re-fitting the GPS model. Afterwards, K-means clustering (KMC)~\cite{KMC} is applied to divide all the subjects into several clusters, where those within the same cluster are similar on one or more GPS components. It is guaranteed that there is at least one subject of each treatment in each cluster. A pair of subjects will be matched if they belong to the same subclass.

\textbf{Ensemble Doubly Robust (enDR)~\cite{enDR}} follows the main idea of DR introduced earlier, and aims to improve the estimation accuracy of both the GPS and ATE. It considers multiple models to estimate the GPS, including multinomial logistic regression, CBPS, random forest, and Generalized Boosted Model (GBM). A rank aggregation technique, with evaluation metric called absolute standardized mean differences, is then applied to determine the optimal GPS estimate. For ATE estimation, enDR also considers multiple potential outcome models, such as linear regression, random forest, and GBM. The key idea of enDR is to ensemble these candidate ATE estimates into an optimal value, and then incorporate it into the doubly robust estimation framework.

\textbf{Generalized Overlap Weight~\cite{overlap}} is constructed as the product of the IPW in Eq. (\ref{eq:ipw}) and the harmonic mean of the GPS in Eq. (\ref{eq:gps}), which is further deduced as:
\begin{equation}
w(t, X)\propto \frac{1/r(t,X)}{\sum_{k=1}^M 1/r(k,X)}.
\end{equation}
This solution corresponds to the target population with the most overlap in covariates across the multiple treatments. 
Furthermore, an empirical sandwich variance estimator~\cite{sandwich} is applied to estimate the causal effects with such generalized overlap weights.

\textbf{Vector-based Kernel Weighting (VBKW)~\cite{VBKW}} is inspired by kernel weights and vector matching. It matches observations with similar propensity score vectors and assigns greater KW to observations with similar probabilities within a given bandwidth:
\begin{alignat}{1}
    w_{i,ATT} &=\left\{
        \begin{aligned}
        1\quad\ &, \forall i \in l_{matched}\\
        k_i(D_{lz}) &, \forall i \in j\\
        \end{aligned}
        \right.   \\ 
    k_i(D_{lz}) &=\left\{
        \begin{aligned}
        \frac{3}{4} \left( 1- \left( \frac{D_{lz}}{h}\right)^2 \right)&, \text{if}\ D_{lz}<h\\
        0\quad\quad\quad&, \text{otherwise}\\
        \end{aligned}
        \right.  \\
    D_{lz} &=|p_i(t=z|x_i)-p_l(t=z|x_i)|.
\end{alignat}
Note that $w_{i,ATT'}$ is constructed in a similar manner to $w_{i,ATT}$. Therefore, $w_{i,ATE}$ can be expressed as $w_{i,ATE} = w_{i,ATT} + w_{i,ATT'}$. The ATE of $z$ vs $z'$ is given as:
\begin{equation}
ATE_{z,z'}=\frac{\sum_{i=1}^n y_i d_i(z)w_{i,ATE}}{\sum_{i=1}^n d_i(z)w_{i,ATE}} - \frac{\sum_{i=1}^n y_i d_i(z')w_{i,ATE}}{\sum_{i=1}^n d_i(z')w_{i,ATE}}.
\end{equation}

\subsubsection{Tree-based Methods} Decision makers are interested about casting which campaign (multi-valued treatment) could obtain the best uplift (ITE or CATE). It can be seen as a map function, i.e. $h(\cdot):\mathbb{X}^d \rightarrow \{1,\dots,m\}$. The goal is to figure out the optimal treatment with the best expected response by $h^*(x)\in \arg\max\limits_{\ t=1,\dots,m} \mathbb{E}[Y|X=x,T=t]$. Tree structure is naturally suitable for this.

\textbf{Contextual Treatment Selection (CTS)~\cite{CTS}} is an example that divides the whole feature space into disjoint subspaces, and each subspace corresponding to a treatment. It gives the probability of a sample falling into any subspace (adopting a treatment) and the corresponding expected response (potential outcome). During the construction of each tree, a recursive binary splitting approach is applied and the goal is to maximize the the expected response.

Splitting criterion is to measure the increase in the holistic expected response $\Delta \mu(s)$ of a candidate split $s$ that divides a leaf node $\phi$ into $\phi_l$ and $\phi_r$. The formal expression is given as follows:
\begin{equation}
\begin{split}
\Delta \mu(s) 
&=P\{X\in\phi_l|X\in\phi\} \max\limits_{t_l=1,\dots,m} \mathbb{E}[Y|X\in\phi_l,T=t_l]\\
&+P\{X\in\phi_r|X\in\phi\} \max\limits_{t_r=1,\dots,m} \mathbb{E}[Y|X\in\phi_r,T=t_r] \\
&-\max\limits_{t_r=1,\dots,m} \mathbb{E}[Y|X\in\phi,T=t].
\end{split}
\end{equation}
In details, $P\{X\in\phi'|X\in\phi\}$ can be regarded as the probability of a subject further falling into $\phi'$ conditioned on already divided to $\phi$. It can be rewritten as $\hat{p}(\phi'|\phi)=\sum_{i=1}^N \mathbb{I}\{x_i\in\phi'\}/\sum_{i=1}^N \mathbb{I}\{x_i\in\phi\}$. Let $\hat{y}_t(\phi')$ denotes the expected response of subspace $\phi'$ given treatment $t$, which is defined as:
\begin{equation}
\hat{y}_t(\phi')=\left\{
\begin{aligned}
\hat{y}_t(\phi)\ \ \ \ \ \ \ \ \ \ \ \ \ \ \ \ \ \ \ \ \ \ \ \ \ \ \ \ \ \ &,\ \text{if}\ n_t(\phi')<\textbf{min\_split}\\
\frac{\left(\sum_{i=1}^ny_i\mathbb{I}\{x_i\in\phi'\}\mathbb{I}\{t_i=t\}+\hat{y}_t(\phi)\cdot\textbf{n\_reg}\right)}{\left(\sum_{i=1}^n\mathbb{I}\{x_i\in\phi'\}\mathbb{I}\{t_i=t\}+\textbf{n\_reg}\right)} &,\ \text{otherwise}\\
\end{aligned}
\right.
\end{equation}
where $\textbf{min\_split}$ is a user-defined parameter, meaning the minimum number of samples required to perform a split. Another parameter $\textbf{n\_reg}$, usually a small positive integer, is provided as a regularity term to avoid misleading from outliers. The response increase can be expressed as:
\begin{equation}
\hat{\Delta \mu}(s)=\hat{p}(\phi_l|\phi)\times \max\limits_{t=1,\dots,M} \hat{y}_t(\phi_l)+\hat{p}(\phi_r|\phi)\times \max\limits_{t=1,\dots,M} \hat{y}_t(\phi_r) - \max\limits_{t=1,\dots,M}\hat{y}_t(\phi).
\end{equation}

Construction of the entire tree is completed when there is no split to conduct according to its termination rules. To alleviate over-fitting of a single tree, CTS creates a forest within which each tree is constructed according to the splitting and termination criteria mentioned above.

\textbf{VAriance Reduced Treatment Selection (VARTS)~\cite{VARTS}} is designed in a similar way as CTS. However, they point out that CTS requires a large amount of training data which is not cost-effective. Therefore, a variance reduced estimator is applied together with the doubly robust estimation technique. Specifically, the expected response is:
\begin{equation}
\hat{V}_{varts}(\phi,t)=\frac{1}{n_\phi}\sum_{i:x_i\in\phi}\left(\frac{\left(Y_i^{obs}-\hat{\mu}_i^{(t)}\right)\mathbb{I}\{T_i=t\}}{p^{(t)}} + \hat{\mu}_i^{(t)} \right).
\end{equation}
Accordingly, the split criterion can be described as:
\begin{equation}
\hat{s}= \arg\max\limits_{s\in\mathcal{S}} \ \hat{p}\left(\phi_l(s)|\phi\right)\times \max\limits_{t_l\in\mathcal{T}}\ \hat{V}_{varts}\left(\phi_l(s),t_l\right)+\hat{p}\left(\phi_r(s)|\phi\right)\times \max\limits_{t_r\in\mathcal{T}}\ \hat{V}_{varts}\left(\phi_r(s),t_r\right).
\end{equation}

\subsubsection{Representation-based Methods}\label{sec:multi-representation}

\textcolor{black}{The CFR framework is extended to solve the problem of multi-valued treatment as well. The most crucial modification lies in how to balance the covariate distributions across multiple groups (corresponding to multiple discrete $T$ values), and how to model the hypothesis function(s) applicable to all these groups.}

An intuitive way to control the confounding bias is using IPM to constrain all the possible pairs of different treatment groups, with a total number of $C_m^2$ for $m$ treatment values. \textbf{Individual Advertising Effect Estimator~\cite{advertising}} proposes a \textit{Transitivity assumption} so that only the IPM of adjacent treatment pairs need taken into account. As for how to design the hypothesis function, all the groups with various treatment assignments share the same network denoted as $f(x,T)=h(\Phi(x),T)$. The objective function is given as follows:
\begin{equation}
\begin{split}
\min\limits_{h,\Phi}\ &\frac{2}{n}\sum_{i=1}^n w_i\cdot L\left(h(\Phi(x_i),t_i),y_i\right)+\lambda\cdot\mathcal{R}(h) +\beta\cdot\sum_{i=1}^{M-1}{\rm IPM_G}(p_\Phi^{T=T_i},p_\Phi^{T=T_{i+1}})\\
&-\frac{\mu_1}{n}\cdot\sum_{i=1}^nL\left(h(\Phi(x_i),t_i),y_i\right)\textbf{1}_{t_i=T_1}\ -\  \dots -\frac{\mu_m}{n}\cdot\sum_{i=1}^nL\left(h(\Phi(x_i),t_i),y_i\right)\textbf{1}_{t_i=T_m} \\
&\text{with}\ \mu_j=\frac{n_j}{n}, w_i=\mu_{t_i},n_j=\sum_{i=1}^n\textbf{1}_{T_j=t_i},j=1,\dots,m.
\end{split}
\end{equation}

\textbf{MEMENTO~\cite{MEMENTO}} follows the intuitive idea that using $C_m^2$ MMD constraints to address the confounding bias, and constructing $m$ hypothesis functions $h_{t_i}$ for counterfactual prediction. The key contribution lies in that it introduces the Expected Precision in Estimation of Heterogeneous Effect (PEHE) loss~\cite{PEHE} to the setting of multi-valued treatment, whose formal definition is:
\begin{equation}\label{eq:PEHE-int}
PEHE_{t,t'}=\int_\mathcal{X}\left(\hat{\tau}_{t,t'}(x)-\tau_{t,t'}(x)\right)^2p(x)\ dx.
\end{equation}
As for the hypothesis network, the prediction loss  given as:
\begin{equation}
\mathcal{L}_F(h)=\sum_tp(t)\int_\mathcal{X}l(x,t)\ p(x|t)\ dx.
\end{equation}
Note that the counterfactual loss can not be directly calculated only using the observational data, an upper bound of $\mathcal{L}_F(h)+\mathcal{L}_{CF}(h)$ is deduced instead:
\begin{equation}
\mathcal{L}_F(h)+\mathcal{L}_{CF}(h)\leq\sum_t\int_\mathcal{X}l(x,t)\ p(x|t)\ dx +C\cdot\sum_t\sum_{t'}(\mu_t+\mu_{t'}){\rm IPM_G}(p(x|t),p(x|t')).
\end{equation}
According to such bound, MEMENTO is designed as an end-to-end model with the goal to minimize the following objective function:
\begin{equation}
\frac{1}{n}\sum_{i=1}^n\frac{1}{\mu_t}\mathcal{L}(y_i,h_{t_i}(\Phi(x_i)))+\mathcal{R}(\Phi,h_{1,\dots,M})\ + \alpha\cdot\sum_{t\neq t'}{\rm MMD}(p(\Phi(x)|t),p(\Phi(x)|t')).
\end{equation}

\subsubsection{Generative Modeling Methods}
Although \textbf{GANITE} is introduced under the setting of binary treatment, it can be easily developed to tackle with multi-valued treatment as well. The key modification lies in the discriminator of the counterfactual block, whose goal is to determine which value of the treatments correspond to the real outcome in $Y_i^F$. Other parts of GANITE remains unchanged. In this way, it can estimate the potential outcome of different treatment values.

\subsubsection{Meta Learning Methods}
\textbf{MetaITE~\cite{meta}} provides a new approach for estimating effects in multi-valued treatments. The key motivation is that the number of samples in different groups is often imbalanced. MetaITE treats a group with sufficient samples as a source domain to train a meta-learner, and then applies gradient descent to update the target domains with fewer samples. There are two core components: 1) A feature extractor $g(\psi):\mathcal{X}\rightarrow\mathcal{Z}$ to obtain balanced embeddings across multiple domains, and 2) An inference network $h(\theta):\mathcal{Z}\rightarrow\mathcal{Y}$ to estimate potential outcomes. During episodic training, MetaITE constructs a support set from a source domain and a query set from a target domain. In the inner loop, the model is optimized on the support set using a loss function designed to ensure the generalization ability across multiple domains:
\begin{equation}
\mathcal{L}_{Sup}=\sum_{i=1}^n\mathcal{L}_{inf}(y_i^{Sup},h(g(X_i^{Sup};\psi);\theta)).
\end{equation}
As for the outer-loop, parameters on the query set will be updated by:
\begin{equation}
\mathcal{L}_{Que}=\sum_{i=1}^n\mathcal{L}_{inf}(y_i^{Que},h(g(X_i^{Que};\psi');\theta')).
\end{equation}
The ultimate goal of training can be concluded as the accumulation of all the losses:
\begin{equation}
\mathcal{L}_{obj}=\mu\mathcal{L}_{Que}+\epsilon\mathcal{L}_{Sup}+\gamma\mathcal{L}_{disc}+\lVert w \rVert_2,
\end{equation}
where $\mathcal{L}_{disc}$ refers to the MMD metric between $g(X^{Sup};\psi$ and $g(X^{Que};\psi)$.

\subsection{With Unobserved Confounders}\label{sec:multi-unobserve}

\textbf{Multiple Causal Estimation via Information (MCEI)~\cite{multiProxy}} follows the idea of recovering a proxy variable $Z$ of the unobserved confounders. Two additional assumptions are proposed in the case of multi-valued treatments. The first is \textit{shared confounding assumption} that the confounders are shared across all treatments, and thus each treatment could reflect some information of the shared confounders. The second is \textit{independence given unobserved confounders}, meaning that treatments are independent given confounders so as to avoid the dependencies between $t_i$ and $t_j$.

A confounder estimator is established to reconstruct a treatment $t_i$ when given the remaining ones $t_{-i}$. Additional Mutual Information (AMI), denoted as $\mathbb{I}(t_i,z|t_{-i})$, is utilized to measure how much additional information could $t_i$ provides for the confounders. Objective function is:
\begin{equation}\label{eq:AMI}
\max\limits_{\theta,\beta}\ \mathbb{E}_{t\sim p(t)}\mathbb{E}_{p_\theta(z|t)}\left[\sum_{i=1}^M\log\ p_\beta(t_i|z)\right]-\alpha\sum_{i=1}^M\mathbb{I}_\theta(t_i,z|t_{-i}),
\end{equation}
where the conditional mutual information can be expressed in the form of conditional entropy. After derivation, the objective function of AMI can be finally updated as:
\begin{equation}
\mathcal{L}=\mathbb{E}_{t\sim p(t)}\mathbb{E}_{p_\theta(z|t)}\left[\sum_{i=1}^M\log\ p_\beta(t_i|z)\right]+\alpha\sum_{i=1}^M\mathbb{G}_{\theta,\xi_i}(z|t_{-i}).
\end{equation}
This lower bound is called multiple causal lower bound (MCLBO)~\cite{multiProxy}, which can be optimized through stochastic gradients by passing the derivative inside expectations.
With the help of \emph{do}-calculus\footnote{Operator $do(t)$ refers to applying intervention on treatment variable $T$ by setting it to a specific value $t$. $T$ and $Y$ are not confounded when $p(y|do(t))=p(y|t)$. Verification of $do(t)$ can be done through simulating the intervention or inferring based on the graph structure.} $do(t^*)$ that eliminates the influence from confounding variables and substitutes the intervention by $t=t^*$ , the formal definition of causal estimation is given below:
\begin{equation}
\mathbb{E}[y|do(t=t^*)]=\mathbb{E}_{p(z)}\mathbb{E}[y|do(t=t^*,z)]=\mathbb{E}_{p(z)}\mathbb{E}[y|t^*,z] .
\end{equation}
In order to learn the outcome model, regression is applied with the residuals and confounder by maximizing the following objective, which can be expressed as:
\begin{equation}
\max\limits_{\eta}\mathbb{E}_{p(y,t)p_\theta(z|t)p(\epsilon_i|z,t)}[\log\ p+\eta(y|z,\epsilon)],
\end{equation}
where $\epsilon_i$ denotes the independent component of the $i$-th treatment. In this way, we have
\begin{equation}
p(y|z,do(t=t^*))=p(y|z,t^*)=p(y|z,\epsilon=d^{-1}(t^*,z)).
\end{equation}

\subsection{Conclusion and Discussion}
Many methods initially proposed for binary treatments can be extended to the multi-valued treatment case. For propensity score, tree-based, and GAN-based methods, the key is to convert the binary problem into a multi-class classification problem. For uplift models using trees, the goal is to determine the best treatment value. GANITE can estimate multi-valued treatment effects by having the discriminator identify which treatment value corresponds to the real data. Extending representation-based approaches is more complex. The key challenges are: 1) learning a shared balanced representation across multiple groups, and 2) modeling hypothesis functions applicable to all groups. There is limited work on addressing unobserved confounders in multi-valued treatments. Attempts have been made to recover proxy variables for the unobserved confounders from treatment assignments, but measuring confounder differences across treatments remains an open challenge.

\section{Continuous Treatment}~\label{sec:continuous}
Another setting included in the complex treatment is that the value of treatment could be continuous. Relevant methods will be introduced from two aspects, covering the methods obeying unconfoundedness assumption and those considering the unobserved confounders.

\subsection{Under Unconfoundedness}\label{sec:continuous-basic}

\begin{figure}[t]
\vspace{-2mm}
	\centering
		\includegraphics[width=0.6\textwidth]{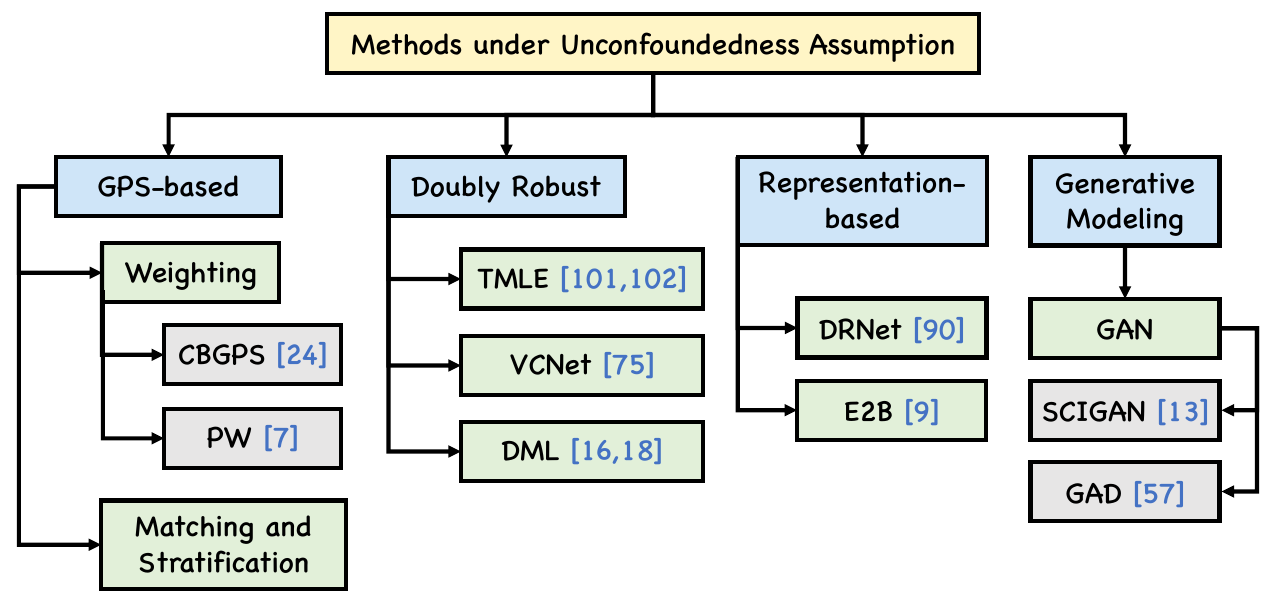}
	\vspace{-2mm}
	\caption{Categorization of continuous treatment methods without unobserved confounders.}
	\label{fig:con-basic}
\end{figure}

As shown in Fig.~\ref{fig:con-basic}, we will introduce four kinds of methods in this section, including GPS-based methods, doubly robust methods, representation-based methods, and generative modeling methods.

\subsubsection{GPS-based Methods}
The GPS mentioned in Eq. (\ref{eq:gps}) is also widely used for the estimation of continuous treatment effect. We discuss the relevant methods utilizing GPS here.

\textbf{Weighting Methods.} Inspired by IPW, the inverse of generalized propensity scoring (IGPS)~\cite{epidemiology} is proposed. To address the problem of possible extreme values in the denominator of the IGPS, researchers also develop a stabilized version called SIGPS~\cite{epidemiology}. For a unit $i$, its SIGPS weight is defined as $w_i^{SIGPS}=\frac{f(T_i)}{r(T_i,X_i)}$, where $f(T_i)$ is the probability density of treatment $T_i$. However, some researchers point out that these methods are sensitive to the model misspecification~\cite{smith2005does,zubizarreta2011matching}. Therefore, optimal balancing weighting is studied, which is concentrated on achieving a direct balance in treatment assignment without the explicit specification of the conditional density $f(T|X)$.

\textbf{Covariate Balancing Generalized Propensity Score (CBGPS)~\cite{moment1}} is a extension of CBPS~\cite{CBPS}, which is mentioned in Section~\ref{sec:binary}, to the setting of continuous treatment. It aims to eliminate the correlation between the treatment $T$ and covariates $X$. In order to ensure the balancing property of the GPS, CBGPS formulates the moment conditions as $\mathbb{E}[w^{CB}TX]=\mathbb{E}[T]\mathbb{E}[T]$, and $\mathbb{E}[w^{CB}]=1$. To maximize its empirical likelihood, the objective can be formally defined as:
\begin{equation}
    \arg\min\limits_{w^{CB}\in\mathcal{W}}\sum_i^n\log w_i^{CB}.
\end{equation}

\textbf{Permutation Weighting (PW)~\cite{permutation-weighting2}} is also introduced as a method for density ratio estimation. For the original data $P$, permutation is performed on $T$ and $X$, resulting in permuted data $Q$. The assignment of $T$ is proved independent of $X$ in $Q$, and $w^{PW}=\frac{p(Q)}{p(P)}$.

\textbf{Matching and Stratification\footnote{In the context of continuous treatment setting, the stratification method can be regarded as a specific instance of the non-bipartite matching method.}.}  The matching method for binary treatment is also generalized to the continuous treatment setting~\cite{matching-continuous}. For example, propensity function~\cite{imai2004causal} is proposed to offer a balancing function, not a one-dimensional score, compared to the GPS. Suppose there is a unique propensity function $\theta(t,x)$ for all $t\in\mathcal{T},x\in\mathcal{X}$, such that $r(t,x)$ depends on $x$ through $\theta(t,x)$. The stratification principle based on the propensity function can be expressed as:
\begin{equation}
    ADRF(t)=\int f\{Y(t)|T=t,\theta(t,x)\}f(\theta(t,x))d\theta(t,x),
\end{equation}
where $f(\cdot)$ refers to the probability density. Researchers also point out that the fundamental concept underlying current matching or stratification methods is discretization. Effectiveness of these methods mainly depends on the choice for distance metrics and the number of strata~\cite{imai2004causal}.

\subsubsection{Doubly Robust Methods} For continuous treatment setting, the purpose of DR model is to predict the average dose-response function via:
\begin{equation}
ADRF(t)=\int_{\mathcal{X}'}\frac{Y-\Psi(t,x')}{r(t,x')}\int_\mathcal{X}r(t,x)dx+\int_\mathcal{X}\Psi(t,x)dx dx'.
\end{equation}

where $r(t,x')$ is the generalized propensity score, $\Psi(t,x)$ is a direct outcome estimator.

\textbf{Targeted Maximum Likelihood Estimation (TMLE)~\cite{TMLE1,TMLE2}} is proposed to alleviate the instability problem for the generalized propensity score. It is developed to estimate the effect of continuous treatment~\cite{DR-con}, rewriting the expectation in the integral form: 
\begin{equation*}
\hat{c}_{h a}(T,X)=\frac{g_{h a}(T) K_{h a}(T)}{r(T,X) / \hat{f}(T)},
\end{equation*}
where $f(T)$ is the probability density, $g_{ha}(T)=(1,(T-a)/h)^T$, $K_{ha}(T)=h^{-1}K\{(T-a)/h\}$, and $K(\cdot)$ is a standard kernel function with bandwidth $h$. By decoupling weighting from the causal inference procedure, TMLE is able to avoid the instability issue from the weighting methods.

\textbf{Targeted Regularizer~\cite{Dragonnet}} can be regarded as a generalization of DR for the binary treatment, predicting the GPS and potential outcome in an end-to-end manner by utilizing neural networks. \textbf{VCNet~\cite{VCNet}} extends such targeted regularization and employs it to the effect estimation of continuous treatment. Its loss function can be expressed as:
\begin{equation}
    \mathcal{L}(\Psi,r,\varphi)=\frac{1}{n}\sum_{i=1}^n\left(Y_i-\Psi(T_i,X_i)-\frac{\sum_{k=1}^K\alpha_k\varphi_k(T_i)}{r(T_i,X_i)}\right),
\end{equation}
where $\varphi_k(\cdot)$ is $k$-th basic function of B-spline, $\alpha$ is a hyper-parameter.

\textbf{Double Machine Learning (DML)~\cite{DML2,DML3}} introduces machine learning techniques (e.g., kernel methods) to achieve double robustness. There is a generalized versions of DML for continuous treatment setting~\cite{DML3}, whose formal definition is:
\begin{equation}
\mathcal{L}(\Psi,r)=\frac{1}{n}\sum_{i=1}^n\Psi(T_i,X_i)+\frac{K_h(T_i-t)}{r(T_i,X_i)}\left(Y_i-\Psi(T_i,X_i)\right),
\end{equation}
where $K_h(T_i-t)$ means the kernel of unit $i$ with treatment $t$, and $h$ refers to the bandwidth of kernel. With such framework, various machine learning techniques can be applied for robust causal inference under the continuous treatment setting.

\subsubsection{Representation-based Methods}
Existing methods mainly focus on the discretization on treatment $T$. \textbf{DRNet~\cite{DRNet}} stratifies the dosage of treatment $T$ into levels for IDRF. It utilizes deep neural network to obtain the representation of covariates $X$ and then learn the representation in each $T$ level. As for ADRF, similar to DRNet,  VCNet discretizes the dosage of $T$ into blocks and learns a decoupled representation $Z$ from $X$. Moreover, VCNet~\cite{VCNet} proposes varying coefficient prediction heads to retain the continuity of dose response curves. \textbf{E2B~\cite{E2B}} is proposed to utilize neural network for the optimization of entropy balancing. It also applies a new training strategy with the objective to directly improve the performance of weighted regression in subsequent for estimating ADRF. CRNet \cite{zhu2024contrastive} propose to learn a double balancing representation via contrastive learning,  aimed at estimating the HDRF while preserving the continuity of treatments. On the basis of conventional MLPs, transformer~\cite{transformer} is introduced to estimate DRFs as well~\cite{transformer-DRF}.

\subsubsection{Generative Modeling Methods}
Inspired by Ganite~\cite{GANITE} that makes the first attempt to introduce GAN technique for the multi-valued setting, \textbf{SciGAN~\cite{SciGAN}} is proposed to estimate IDRF for continuous treatment. Two hierarchical discriminators are applied, where one is to distinguish the type of treatment and the other is to tell from the exact dosage of it. \textbf{Generative Adversarial De-confounding (GAD)~\cite{GAN-ADRF}} algorithm using GAN to learn the IPW of treatment for ADRF follows the idea of permutation weighting~\cite{permutation-weighting2}. It models the distribution of propensity score implicitly via GAN which deems permuted distribution $P(T|X)$ as the ground truth and puts it into the discriminator with data from the generator.

\subsection{With Unobserved Confounders}\label{sec:continuous-unobserve} 
Similar to the methods focused on binary treatment, proxy and IV are still the main paradigms used for continuous treatment when considering the presence of unobserved confounders. 

\begin{figure}[t]
	\centering
		\includegraphics[width=1\textwidth]{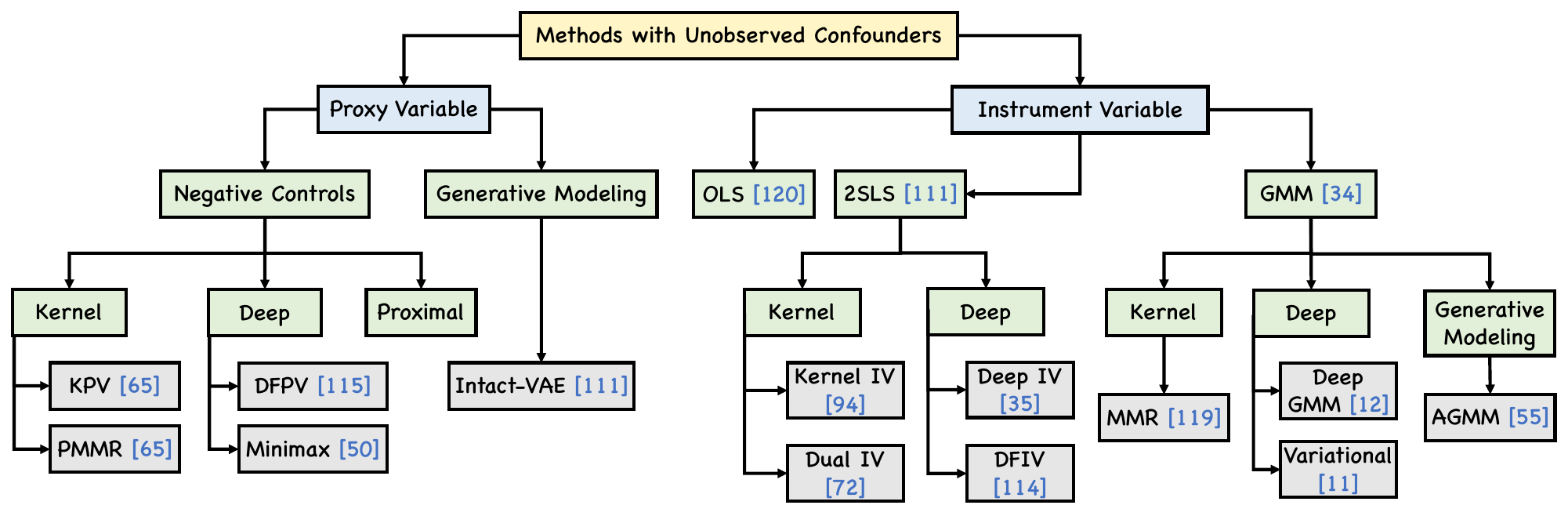}
        \vspace{-5mm}
	\caption{Categorization of continuous treatment methods with unobserved confounders.}
	\label{fig:con-bunobs}
    \vspace{-5mm}
\end{figure}

\subsubsection{Proxy Variable} Main idea of proxy has been introduced in Section~\ref{sec:binary}, and we give a further classification of the relevant methods into kernel methods, deep methods, and proximal methods.

\textbf{Kernel Methods.}
A kernel based negative control method~\cite{singh2020kernel} is generalized from the kernel IV method~\cite{singh2019kernel}. Given $\phi(h)$ as the feature map from $h$ to RKHS and $k_h$ as the kernel function of $h$, the RKHS regularity conditions are assumed as follows: (1) $k_T$, $k_X$, $k_O$, and $k_E$ are continuous and bounded. (2) $k_X$ and $k_O$ are characteristic. (3) $\phi(T)$, $\phi(X)$, $\phi(O)$, and $\phi(E)$ are measurable. Suppose $h_0\in H$, the ADRF under the continuous treatment setting can be derived as:
\begin{equation}
ADRC(t)= b(t,\mu), \mathcal\mu=\int[\phi(X) \otimes \phi(O)] \mathrm{d} f(X, O).
\end{equation}

Two-Stage Least Square (2SLS) and Maximum Moment Restriction (MMR) methods are classic methods in IV. Both of them can be extended to the proxy paradigm, and the new methods are called \textbf{Kernel Proxy Variable (KPV)~\cite{mastouri2021proximal}} and \textbf{Proxy Maximum Moment Restriction (PMMR)~\cite{mastouri2021proximal}}, respectively. The objective of them is to minimize $\mathbb{E}[Y-G(T,E,X)]^2$, where $G(t,e,x)=\int_\mathcal{O}f(t,x,o)g(o|t,x,e)do$.

\textbf{Deep Methods.}
\textbf{Deep Feature Proxy Variable (DFPV)~\cite{DFPV}} introduces a representation-based method to the proxy framework. It approximates the bridge function of proxy via a 2SLS method. The objective of the first stage in 2SLS with proxy variables is a conditional mean embedding which is localized as the distribution $P(O|T,X,E)$. For the second stage in 2SLS with the proxy, the interested estimates are regressed via empirical risk minimization (ERM). 

A \textbf{minimax estimator~\cite{minimax}} applying GMM to estimate $ADRF(t)=\int Y(t)\pi(t|x)d\mu(u)$, where $\mu(\cdot)$ is a Lebesgue measure. In addition, $\pi$ is a contrast function and $\pi(T|X)$ could be modeled as GPS. The moment conditions are given as follows:
\begin{equation}
\mathbb{E}[Y-h_0(O,T,X)|E,T,X]=0
\end{equation}
\begin{equation}
\mathbb{E}\left[\pi(T|X)\left(q_0(E,T,X)-\frac{1}{f(T|O,X)}\right)|O,T,X\right]=0,
\end{equation}
where $h_0(\cdot)$ and $q_0(\cdot)$ are square-integrable functions. Note that this minimax estimator can be implemented via kernel methods or representation-based methods.

\textbf{Proximal Methods.} A sieve method~\cite{deaner2018proxy} utilizing Penalized Sieve Minimum Distance (PSMD) is proposed to estimate causal effect for continuous treatment. Besides, the researchers reduce the causal effect estimation problem to a linearity setting and designs a doubly robust method~\cite{deaner2021proxy} for the linear model with proxy.

\textbf{Generative Modeling Methods.} Inspired by CEVAE in the binary treatment setting, \textbf{Intact-VAE~\cite{Intact-VAE}} generalizes the prognostic score to estimate DRFs under the circumstance of continuous treatment. NC methods is proposed in recent years, and there is seminal literature to expound on the theory for its identification. However, few works support the complete theory of proxy methods with VAEs. To conclude, it is still an challenge that remains more exploration for the identification and estimation with proxy via deep latent variable methods.

\subsubsection{Instrumental Variable (IV)}\label{sec:continuous-IV}
\textcolor{black}{The brief introduction of IV can be referred to Section~\ref{sec:binary}. Although it is initially proposed to address the binary treatment effect estimation, it can be naturally developed into the continuous treatment setting.}

\textbf{Ordinary Least Square (OLS)~\cite{OLS}} is a classical approximation method for the linear model, which is constructed as follows:
\begin{equation}\label{eq:nonlinear}
T=\alpha Z+\epsilon_1, Y=\beta T +\epsilon_2,
\end{equation}
where $\epsilon_1$ and $\epsilon_2$ are error terms, and $\mathbb{E}[\epsilon_2|Z]=0$. In this way, the outcome $Y$ could be directly regressed on $Z$. However, the drawback of OLS lies in that the estimate is proved to be biased~\cite{angrist_does_1991}.

\textbf{Two-Stage Least Square (2SLS)~\cite{2SLS}} can be regarded as an extension of OLS, where the data generation paradigm is modeled as nonlinear but additive, i.e. $T=f(Z)+\epsilon_1, Y=g(T)+\epsilon_2$. The first stage in 2SLS is to regress $T$ on $Z$, thus obtaining the fitted treatment $\hat{T}$. As for the second stage, the outcome $Y$ is regressed on $\hat{T}$. Researchers then generalize the linear functions to nonlinear setting~\cite{amemiya_nonlinear_1974}, where the objective of the interested parameter $\beta$ is designed as:
\begin{equation}
\arg\min\limits_{\beta}(Y-\hat{Y})'f(Y-\hat{Y}).
\end{equation}
Note that $g(\cdot)$ is a function of instrument $Z$ and unobserved confounders $U$, and how to model $g(\cdot)$ in the non-parametric case becomes a new challenge. A solution is to consider it an ill-posed inverse problem~\cite{newey_instrumental_2003}, i.e. rewriting Eq. (\ref{eq:nonlinear}) in the integral form:
\begin{equation}\label{eq:ill-posed}
\mathbb{E}[Y|Z]=\mathbb{E}[g(T)|Z]=\int g(T,Z)F(dT|Z),
\end{equation}
where $F(dT|Z)$ is the conditional cumulative distribution function of $T$ given $Z$. We can then consider the non-parametric modeling $g(\cdot)$ as solving a Fredholm integral equation. Feasible solutions include kernel methods, deep methods, and etc.

\textbf{Kernel Methods.}
These methods are widely applied in non-parametric IV estimation, where the kernel functions is to approximate the conditional density of $Y$ given $T$ and $Z$. Following the structure function in Eq. (\ref{eq:nonlinear}), the joint probability density of $T$ and $Z$ can be defined as follows:
\begin{equation}
\frac{1}{\sigma^2}\sum_{j=1}^nK(T_i-T_j,T_i)K(Z_i-T_j,Z_i),
\end{equation}
where $K(\cdot,\cdot)$ refers to the generalized kernel estimator. When it comes to the non-additive setting modeled as $Y=g(T,Z,\epsilon)$. The SEM is generalized~\cite{matzkin_nonparametric_2003} with the join probability density function defined as $g(Y,W)=\frac{1}{n_\sigma}\sum_{j=1}^nK\left(\frac{Y_i-Y_j}{\sigma},\frac{W_i-W_j}{\sigma}\right)$, where $\sigma$ is the bandwidth of kernel function $K(\cdot,\cdot)$ and $W=(T,Z)$. In addition, the conditional density of $Y$ given $W$ is formulated as $g_{Y|W}(Y)=\frac{g(Y,W)}{\int_{-\infty }^{\infty}g(\epsilon,W)}$. The non-additive assumption could be implemented in a different way~\cite{altonji_cross_2005}, i.e. defining the joint probability density as:
\begin{equation}
\frac{1}{\sigma}\sum_{j=1}^nK\left(\frac{Y_i-Y_j}{Y_i}\right)K\left(\frac{T_i-T_j}{T_i}\right)K\left(\frac{Z_i-Z_j}{Z_i}\right).
\end{equation}
\textbf{Kernel IV~\cite{singh2019kernel}} is also proposed which optimizes a conditional means mapping $m:\mathcal{H}_\mathcal{T}\rightarrow\mathcal{H}_\mathcal{Z}$, where $\mathcal{H}_\mathcal{T}$ and $\mathcal{H}_\mathcal{Z}$ are scalar-valued RKHSs, and $m':\mathcal{H}_\mathcal{Z}\rightarrow\mathcal{H}_\mathcal{T}$. The objective of kernel IV is:
\begin{equation}
W=K_{XX}(K_{ZZ}+n\lambda I)^{-1}K_{Z\tilde{Z}},
\end{equation}
where $K_{XX}$ and $K_{ZZ}$ are the empirical kernel matrices. The ill-posed inverse problem can be converted to the convex-concave saddle-point problem, and \textbf{Dual IV~\cite{DualIV}} is thus proposed avoiding the regression in the first stage of 2SLS. Objective function of Dual IV can be formally defined as $\min_{f\in\mathcal{F}}\max_{g\in\mathcal{G}}\mathbb{E}_{T, Y, Z}[f(T)g(Y,Z)]-\mathbb{E}_{Y, Z}[l(g(Y,Z))]$
, where $l(\cdot)$ is a loss function, $\mathcal{F}$ is the space of functions over $T$, and $\mathcal{G}$ is the space of function over $Y$ and $Z$.

\textbf{Deep Methods.}
Deep models have been widely used in IV methods as a powerful modeling tool. As for 2SLS, \textbf{DeepIV~\cite{DeepIV}} is a good example that applies a neural network to the two-stage method. It models the conditional distribution of treatment given instrument and treatment $P(T|Z,X)$ in the first stage and approximates the causal effects in the second stage. In another way, the neural network could be directly introduced to linear 2SLS as~\textbf{DFIV~\cite{xu_learning_2020}}, which encodes representation into a linear space. Afterwards, the representation can be directly thrown into the linear 2SLS.

\textbf{Generalized Method of Moments (GMM)~\cite{GMM1,GMM2}} is a such an efficient one-stage method with semi-parametric estimation, which could be seen as an extension of 2LSL. It constructs moment restrictions based on the characteristics of IV and directly estimates the structure function of $T$. There are also some variations of GMM, and we classify them into three categories as: kernel methods, deep methods, and generative modeling methods. \textbf{Maximum Moment Restriction (MMR)~\cite{MMR}} is such an example of kernel methods. As for the deep methods, \textbf{Deep GMM~\cite{deepGMM}} introduces a neural network to GMM and can be applied to image data. It can be further generalized to a min-max game formulation~\cite{liao_provably_2020} which constructs SEM with the proposed representation-based method. There is also a \textbf{variational method~\cite{bennett_variational_2020}} of moments that provides a new way for representation-based implementation.

\textbf{Generative Modeling Methods.} GAN is also employed to construct conditions of GMM, and this model is called \textbf{AGMM~\cite{lewis_adversarial_2018}}. The objective function is to measure the discrepancy between the observed moments and the moments implied by the model. The generator generates synthetic data based on the estimated parameters, while the discriminator tries to distinguish between the real observed data and the synthetic data generated by the generator. By combining moment conditions with adversarial training, AGMM is able to handle high-dimensional datasets. It also takes into account potential distributional biases, providing more robust parameter estimates.

\subsection{Conclusion and Discussion}
Effect estimation of the continuous treatment can be regarded as extension or generalization of that of binary or multi-valued treatment. GPS is also widely used in the setting of continuous treatment, and the corresponding weighting, matching, and doubly robust approaches are naturally developed as well. Concerning about the misspecification of GPS, there are many other attempts with representation-based methods and GAN-based methods. The key modification for representation-based methods still lies in learning the balanced representation for continuous treatment so as to decorrelate $T$ with $X$. As for the approaches using GAN, limitation is the lack of theoretical guarantees. As for the methods tackling with unobserved confounders, proxy and IV approaches can be naturally utilized. Challenges of them have been summarized in Section~\ref{sec:binary}.

\section{Bundle Treatment}~\label{sec:bundle}
Different from the setting of multi-valued treatment $T\in\mathcal{T}^{mul}\subset \mathbb{R}$, a unit can simultaneously adopt several treatments $T\in\mathcal{T}^{bun}\subset\{0,1\}^m$ in the case of bundle treatment. Studies on this setting can also be divided into two categories, according to the existence of unobserved confounders.

\begin{figure}[t]
	\centering
		\includegraphics[width=0.6\textwidth]{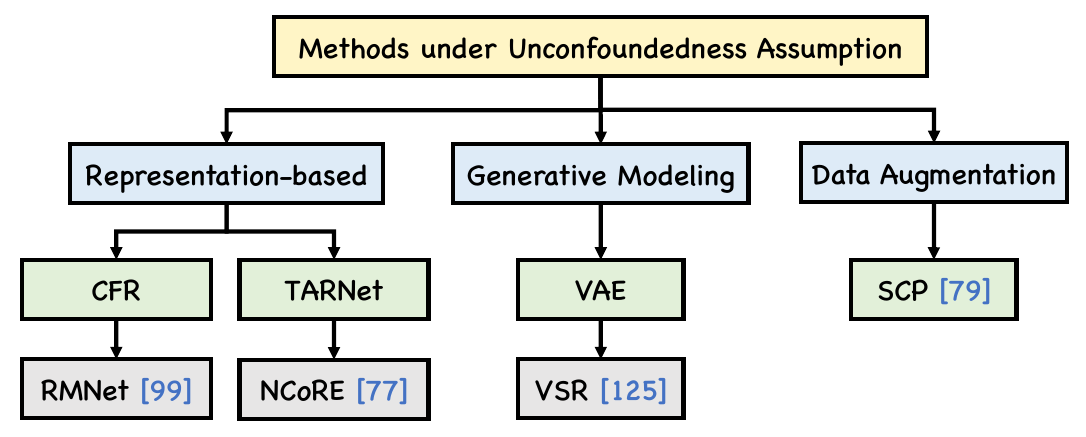}
	\vspace{-2mm}
	\caption{Categorization of bundle treatment methods without unobserved confounders.}
	\label{fig:bundle-basic}
    \vspace{-2mm}
\end{figure}

\subsection{Under Unconfoundedness}\label{sec:bundle-basic}
Methods related to bundle treatment and in accordance with the unconfoundedness assumption are included in Fig.~\ref{fig:bundle-basic}. The majority of them belong to the representation-based solutions or generative modeling methods. Single-cause perturbation method explores a new way for estimating the simultaneous intervention on multiple variables by the means of data augmentation. We will give a brief introduction of these methods in this section. 

\subsubsection{Representation-based Methods}

\textcolor{black}{CFR or TARNet are also developed into the setting of bundle treatment. One challenge is the more complex confounding bias, since the possible treatment space expands in an exponential manner, and a shared hypothesis network is thus needed for all the treatment groups with the purpose of sample efficiency. Moreover, the additional influence caused by interactions among treatments taken in the same time should also be considered.}

\textbf{Regret Minimization Network (RMNet)~\cite{Regret}} is  proposed to address the problem of sample efficiency, together with the gap between the regression accuracy for the whole treatment space and the decision-making performance with respect to an exact treatment regime.

Decision-focused risk is proposed to mitigate such gap mentioned above, which is essentially a classification task to predict whether a treatment is assuredly better in term of the decision-making performance. The goal is to minimize the following comparison loss, i.e.
\begin{equation}
{\rm ER}_\mu^u(f)=\mathop{\mathbb{E}}\limits_{x}\left[\frac{1}{|\mathcal{T}|}\sum_{t\in\mathcal{T}}I(y_t\geq\bar{y}\oplus f(x,a)\geq\bar{y})\right],
\end{equation}
where $\oplus$ denotes the the logical XOR and $\bar{y}=\mathbb{E}_{t\in\mu(t|x)}[Y_t|x]$ is the average performance of past decision-makers under $x$. Replacing $\bar{y}$ with $g(x)\simeq\mathbb{E}_{t\in\mu(t|x)}[Y_t|x]$ and substituting the $0$-$1$ loss with cross-entropy gives:
\begin{equation}
\begin{split}
\widetilde{\rm ER}_g^u(f)&=\mathop{\mathbb{E}}\limits_{x}\left[-\frac{1}{|\mathcal{T}|}\sum_{t\in\mathcal{T}}\{s\log v+(1-s)\log(1-v)\}\right] \\
{\rm with}\ s&:=I(y-g(x)\geq0),\ v:=\sigma(f(x,t)-g(x)).
\end{split}
\end{equation}
Considering that the regression error (MSE) still plays an important role on decision-making, loss function with respect to the hypothesis network $h(\Phi)$ is formally defined as:
\begin{equation}\label{eq:regret}
\mathcal{L}^u(f;g)=\sqrt{\ \widetilde{\rm ER}_g^u(f)\cdot{\rm MSE}^u(f)}\ .
\end{equation}

As for the challenge of sample efficiency in the representation network $\Phi$, embeddings of $x$ and $t$ are both learned for the following inference. There are two alternative plans, and the first is to construct a single network $\Phi$ to learn the joint representation and utilize IPM as a restriction:
\begin{equation}
{\rm IPM_G}(p_1,p_2)=\mathop{\rm sup}\limits_{g\in G}\left|\int_\mathcal{S}g(s)(p_1(s)-p_2(s))\ ds\right|.
\end{equation}
The second method is to construct two separate networks $\Phi_x$ and $\Phi_t$ and regularize them to be independent from each other by minimizing the Hilbert-Schmidt Independence Criterion (HSIC)~\cite{hsic}: 
\begin{equation}
{\rm HSIC}(p(\phi_x),p(\phi_t))={\rm MMD}^2(p(\phi_x,\phi_t),p(\phi_x)p(\phi_t)).
\end{equation}
Finally, the objective function can be concluded as:
\begin{equation}
\min\limits_{f}\ \frac{1}{n}\sum_{i=1}^n\mathcal{L}(f(x_i,t_i),y_i;g(x_i),\beta_i)+\alpha\cdot D_{bal}(\{\phi(x_i,t_i)\}_i)+\mathcal{R}(f),
\end{equation}
where $\mathcal{L}$ is the empirical instance-wise version of Eq. (\ref{eq:regret}), and $D_{bal}$ is the balancing regularizer (IPM or HSIC) corresponding to the design of the representation network. However, one of the key challenges of causal inference with bundle treatment is to explicitly measure the mutual influence among multiple treatments that are adopted at the same time.

\textbf{Neural Counterfactual Relation Estimation (NCoRE)~\cite{NCoRE}} makes attempt to model such cross-treatment interactions by analogizing their superimposed effects to the additive mechanism of layers in the neural network. NCoRE refers to the design of TARNet~\cite{CFR} that NCoRE constructs an arm for each treatment just like the the multiple heads of $h(\Phi)$ in TARNet. The difference is that there is an merge layer connecting all the treatment arms in the end. When training the model, all the samples with the information of $x$ will go through the base layers, while only the arms corresponding to those treatments involved in the bundle treatment will be updated. In the stage of prediction, the merge layer receives the outputs from related treatment arms as inputs and finally calculate the potential outcome.

\subsubsection{Generative Modeling Methods} In the setting that bundle treatment can be regarded as a high-dimensional vector, \textbf{Variational Sample Re-weighting (VSR)~\cite{VSR}} points out that it is feasible to learn a latent representation $Z$ for $T$ using VAE, and then decorrelate the low-dimensional $Z$ with confounders $X$. Specifically, the objective function of VAE is to maximize the following Evidence Lower Bound $\mathcal{L}_{ELBO}$:
\begin{equation}
\frac{1}{n}\sum_{i=1}^n\mathbb{E}_{z\sim q_\phi(z|t_i)}\left[\log_{avrphi(t_i|z)}+\log p(z)-\log q_\phi(z|t_i)  \right]
\end{equation}
To remove the confounding bias, variational sample weight $w^d=\{w_i^d\}_{i=1}^n$ is also proposed as:
\begin{equation}
w_i^d=W_T(x_i,t_i)=\frac{p(t_i)}{p(t_i|x_i)}=\frac{1}{\mathbb{E}_{z\sim q_\phi(z|t_i)}\left[\frac{1}{W_Z(x_i,z)}\right]},
\end{equation}
where $W_Z(X,Z)$ is the density ratio estimation with which $T$ can be decorrelated with $X$. Specifically, the data points from observational dataset are regarded as positive samples ($L=1$) while those from decorrelated target dataset are negative samples ($L=0$). In this way, we define:
\begin{equation}
W_Z(X,Z)=\frac{p(X,Z|L=0)}{p(X,Z|L=1)}=\frac{p(L=1)}{p(L=0)}\cdot\frac{p(L=0|X,Z)}{p(L=1|X,Z)}=\frac{p(L=0|X,Z)}{p(L=1|X,Z)}.
\end{equation}
There is a classifier $p_{\theta_d}(L|X,Z)$ to give the values of $p(L|X,Z)$ with the limitation that $\frac{p(L=1)}{p(L=0)}=1$ for all the data points. Finally, a network $f_{\theta_p}(x_i,t_i)$ is learned to predict the potential outcome, and the entire loss function can be concluded as below:
\begin{equation}
\mathcal{L}_{pre}=\frac{1}{n}\sum_{i=1}^nw_i^d\cdot\mathcal{L}\left(f_{\theta_p}(x_i,t_i),y_i\right).
\end{equation}

\subsubsection{Data Augmentation}
\textbf{Single-cause Perturbation (SCP)~\cite{SCP}} provides a new insight that the accuracy of causal inference for bundle treatment can be improved by the means of data augmentation on counterfactual predictions. In other words, SCP perturbs a single treatment ($m$ in total) to be its opposite value, and thus generates an additional dataset by predicting the potential outcomes. In this way, the treatment assignment becomes more balanced than the original observational data, which mitigates the confounding bias in an entirely new manner.

Under the \textit{sequential ignorability assumption}~\cite{SeqIgn}, conditional expectation of the potential outcome for a single treatment has equivalence to that of a bundle treatment, i.e.
\begin{equation}
\mathbb{E}\left[Y(t_m,t_{-m})|X\right]=\mathbb{E}\left[Y(t_m)|X,T_{-m}(t_m)=t_{-m}\right],
\end{equation}
where $t_m$ is a single treatment, namely the $i$-th treatment in the bundle, while $t_{-m}$ means the left ones in the bundle treatment. \textcolor{black}{This assumption plays a key role for the validity of SCP, simplifying the problem of treatment effect estimation for bundle treatment into that for a single treatment.}

Three modules are included in the design of algorithm, including single-cause model training, data augmentation, and covariate adjustment. The goal of the first module is to well estimate the holistic potential outcome $\mathbb{E}[Y|X'_m,T^\downarrow_{-m},T_m]$, where two estimators are need for a single treatment $t_m$ and its causal descendants $T^\downarrow_{-m}(t_m)$, respectively. Disentangled Representations for Counterfactual Regression algorithm (DR-CFR)~\cite{DR-CFR} is applied here as the estimators. It is able to sample perturbed data points for the data augmentation once the single-cause model is fitted, the process of which can be briefly summarized as $3$ steps:
\begin{enumerate}
    \item From the observational data $(x,y,t)\in\mathcal{D}_0$, directly obtaining the non-descendants of $t'_m$, which are denoted as $t^\uparrow_{-m}(t'_m)$. This is because non-descendants are unaffected by the intervention, i.e. $t^\uparrow_{-m}=t^\uparrow_{-m}(0)=t^\uparrow_{-m}(1)$.
    \item Sampling $t^\uparrow_{-m}(t^\prime_m)\sim P(T^\uparrow_{-m}(t^\prime_m)|X^\prime_m)$.
    \item Calculating $y(t'_m):=\mathbb{E}(Y(t'_m)|X'_m,T_{-m}(t'_m))$.
\end{enumerate}
Note that $t'_m=1-t_m$ refers to perturbing treatment $T_m$ where $t_m\in\{0,1\}$. To generate a new data point $(x,\tilde{y}^k,\tilde{t}^k)$, denote $\tilde{y}^k:=y(t'_m)$ and $\tilde{t}^k:=(t'_m,t_{-m}(t'_m))$. The perturbed dataset is $\mathcal{D}_{m}=\{x_i,\tilde{y}_i^m,\tilde{t}_i^m\}_{i=1}^n$. The newly generated data for all the single treatments are merged together as the augmented dataset. As for the covariate adjustment, a standard feed-forward neural network $f_\theta:\mathbb{R}^D\times\Omega\rightarrow\mathbb{R}$ is trained on the augmented dataset to learn the following conditional expectation:
\begin{equation}
\mathbb{E}[Y(t)|X=x]=\mathbb{E}[Y|X=x,T=t].
\end{equation}

\subsection{With Unobserved Confounders}\label{sec:bundle-unobserve}
\begin{figure}[t]
\vspace{-3mm}
	\centering
		\includegraphics[width=0.68\textwidth]{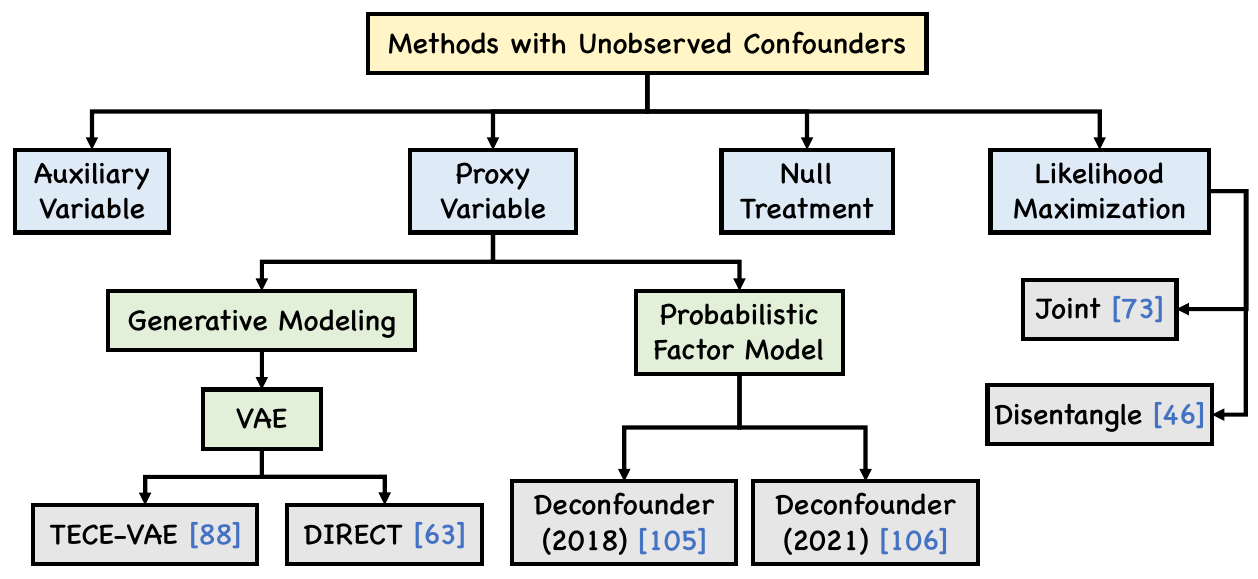}
	\vspace{-3mm}
	\caption{Categorization of bundle treatment methods with unobserved confounders.}
	\label{fig:bundle-unobs}
\end{figure}

The problems of unobserved confounders have also been studied under the circumstance of bundle treatment. Relevant methods are recorded in Fig.~\ref{fig:bundle-unobs}.

\subsubsection{Proxy Variable.}

\textbf{TECE-VAE~\cite{TECE-VAE}} expands CEVAE to the setting of bundle treatment, and its contribution lies in introducing task embedding to model the interdependence among multiple treatments. It allows a flexible representation of a task by multiplying a vector of zeros and ones, meaning which treatments are applied, and a weight matrix $W$ is learned. 

An encoder and a decoder is included in TECE-VAE. As for the former, a network $g_1$ is trained for the distribution $q(t|x)$ given $x$ as the input, from which the treatment vector $\tilde{t}$ is sampled. Multiplying $\tilde{t}$ with the embedding matrix $W$ gives the new representation $\tau=W\tilde{t}$. Afterwards, another network $g_2$ is trained for the distribution $q(y|t,x)$ given $\tau$, where the potential outcome $\tilde{y}$ is sampled. Combining $\tau$, $x$, and $\tilde{y}$ as the input, networks $g_3$ and $g_4$ output the mean value and variance of $q(z|t,x,y)$, respectively.

The purpose of the decoder is to well reconstruct $x$, $t$ and $y$, with the input $z$ sampled from $q(z|t,x,y)$ mentioned above. Networks $f_1\sim f_4$ are established in case of the various data form of $x$ (binary, categorical, or continuous). Design of $f_5$ is similar to $g_1$, whose output is the distribution of $q(t|z)$ for sampling the treatment vector $\tilde{t}$. The new representation $\tau$ is obtained in the same way as described in the encoder. Ultimately, $f_6$ aims to learn the distribution of $p(y|t,z)$ for the determination of $y$ given $z$ and $\tau$.

\textbf{Disentangled Multiple Treatment Effect Estimation (DIRECT)~\cite{DIRECT}} aims to learning the representation of confounder proxy $Z$ from the treatment assignments by VAE, and further explores the interdependence of multiple treatments. There are two main blocks including an inference network and a generation network. The objective of the inference network is to learn the disentangled representation of $Z$. To be specific, the embeddings of every single treatment $t_j$ are learned according to the treatment assignment $A$, followed by a clustering module $f_c(\cdot)$ to approximately simulate the distribution of each class, i.e. $q(C_j|t_j)=Mult\left(f_c(t_j)\right)$. Notation $C_j$ here represents the cluster assignment of $t_j$, and $Mult(\cdot)$ is Multinomial distribution. Such idea is similar to VaDE~\cite{VaDE}. Afterwards, disentangled confounder representation $Z^{(k)}$ is learned for each class, which is implemented in a manner similar to $\beta$-VAE~\cite{beta-VAE}. The holistic confounder representation of $Z$ is obtained by concatenating all of them. 

In the generation network, the main task is to reconstruct the treatment assignment $A$ when given the representation $T$ along with $Z$. Moreover, the observational outcomes are also used as supervision for better capturing the latent confounders, and the prediction loss is defined as:
\begin{equation}
\mathcal{L}_y=-\sum_{i=1}^n\log p(\hat{Y}_i=y_i|z_i,a_i,T).
\end{equation}
Following the classic VAE scheme, the loss function of DIRECT can be derived as:
\begin{equation}
\begin{split}
\mathcal{L}=&-\mathbb{E}[\log p(A|Z,T,C)]+\mathbb{E}_{q(T|A)}KL\left(q(C|T)\Vert p(C)\right) \\
&+\mathbb{E}_{q(C|T)}KL\left(q(T|A)\Vert p(T|C)\right) + \lambda\mathcal{L}_y \\
&+\beta\sum_{k=1}^K\mathbb{E}_{q(T|A)q(C|T)}KL\left(q\left(Z^{(k)}|A\right)\Vert p\left(Z^{(k)}\right)\right).
\end{split}
\end{equation}
Hyper-parameters $\beta$ and $\lambda$ are used to control the effect of different parts.

Another way to capture the latent confounders of multiple treatments is using probabilistic factor model. \textbf{Deconfounder (2018)~\cite{deconfounder-2018}} is proposed out of concern that a variable making all the treatments conditionally independent from each other could be found, once a factor model well representing the treatment distribution is figured out. Details of implementations can be concluded as the following steps.
\begin{enumerate}
    \item Finding out a suitable model for latent variable according to the treatment assignment, namely fitting a probabilistic factor model to capture the joint distribution among them:
    \begin{equation}
        \left\{
        \begin{aligned}
          &\ Z_i\sim p(\cdot|\alpha)\quad i=1,\dots,n \\
          &\ T_{ij}|Z_i\sim p(\cdot|z_i,\theta_j)\quad j=1,\dots,m
        \end{aligned}
        \right.
    \end{equation}
    where $\alpha$ refers to the parameters for distribution of $Z_i$, and $\theta_j$ denotes those for the per-cause distribution of $T_{ij}$. Note that $i$ is the index for each sample while $j$ is that of each cause.
    \item Inferring the latent variable for each sample:
    \begin{equation}
        \hat{z}_i=\mathbb{E}_M[Z_i|T_i=t_i].
    \end{equation}

    \item Estimating the casual effect by utilizing $\hat{z}_i$ as a substitute of the confounders:
    \begin{equation}
        \mathbb{E}[Y_i(t)]=\mathbb{E}\left[\mathbb{E}\left[Y_i(t)|\hat{Z}_i,T_i=t\right]\right].
    \end{equation}
\end{enumerate}

\textbf{Deconfounder (2021)~\cite{deconfounder-2021}} is a improved version of Deconfounder (2018) by the same authors. There are some key findings that a subset $\mathcal{C}$ of treatments could be regarded as proxies of the unobserved confounders so as to help the causal identification for the remaining treatments. The distribution of $\mathcal{C}$ can be determined as well. Implementations are similar to those of Deconfounder (2018), which are described as follows.
\begin{enumerate}
    \item Constructing latent variable $\hat{Z}$ that makes all the treatment conditionally independent with each other, i.e.
    \begin{equation}
        \hat{P}(t_1,\dots,t_m,\hat{z})=\hat{P}(\hat{z})\prod_{j=1}^m\hat{P}(t_j|\hat{z}),
    \end{equation}
    where $\hat{P}(\cdot)$ is consistent with the observational data that $P(t_1,\dots,t_m)=\int\hat{P}(t_1,\dots,t_m,\hat{z})d\hat{z}$.

    \item Fitting the outcome model $P(y,t_1,\dots,t_m)$ by $\int\hat{P}(y|t_1,\dots,t_m,\hat{z})\hat{P}(t_1,\dots,t_m,\hat{z})\ d\hat{z}$.

    \item Estimating the treatment distribution:
    \begin{equation}
        \hat{P}\left(y|{\rm do}(t_\mathcal{C})\right)\triangleq\int\hat{P}(t_1,\dots,t_m,\hat{z})\times\hat{P}(t_{\{1,\dots,m\}\backslash\mathcal{C}},\hat{z})\ d\hat{z}dt_{\{1,\dots,m\}\backslash\mathcal{C}}.
    \end{equation}
\end{enumerate}

\subsubsection{Auxiliary Variable.}
Similar to confounder proxy and IV, auxiliary variable~\cite{miao2022identifying} has no causal relationship with the outcome according to the following assumption. The first is \textit{exclusion restriction}, i.e. $Z\perp \!\!\! \perp Y|(X,U)$. Two additional assumptions are proposed to limit the joint distribution between treatments and the unobserved confounders:
\begin{enumerate}
    \item \textit{Equivalence}. For any $\alpha$, any $\tilde{f}(x,u|z)$ that solves $f(x|z;\alpha)=\int_u\tilde{f}(x,u|z)du$ can be written as $\tilde{f}(x,u|z)=f\{X=x,V(U)=u|z;\alpha\}$ for an invertible but not necessarily known function $V$.
    \item \textit{Completeness}. For any $\alpha$, $f(u|x,z;\alpha)$ is complete in $z$, i.e. for any fixed $x$ and square-integrable function $g$, $E\{g(U)|X=x,Z;\alpha\}=0$ almost surely if and only if $g(U)=0$ almost surely.
\end{enumerate}

Equivalence is a high-level assumption stating that the treatment-confounder distribution lies in a model that is identified upon a one-to-one transformation of $U$. Because the unconfoundedness assumption holds conditional on any one-to-one transformation of $U$, this allows us to use an arbitrary admissible treatment-confounder distribution to identify the treatment effects. 

Completeness is a fundamental concept in statistics, meaning that conditional on $X$, any variability in $U$ is captured by variability in $Z$, analogous to the relevance condition in the instrumental variable identification. When both $U$ and $Z$ have $k$ levels, completeness means that the matrix $[f(u_i|x,z_j)]_{k\times k}$ consisting of the conditional probabilities is invertible.

Under these assumptions, identification with auxiliary variable is proved feasible. Algorithm is described as follows.
\begin{enumerate}
    \item Obtaining an arbitrary admissible joint distribution $\tilde{f}(x,u,z)$.
    \item Using the estimate from Step (1), along with an estimate of $f(y|x,z)$, to solve the following equation for $\bar{f}(y|u,x)$:
    \begin{equation}
        f(y|x,z)=\int_u\tilde{f}(y|u,x)\tilde{f}(u|x,z)du.
    \end{equation}

    \item Plugging the estimate of $\tilde{f}(y|u,x)$ from Step (2) and the estimate of $\tilde{f}(u)$ derived from $\tilde{f}(u,x,z)$ into the equation below to estimate $f\{Y(x)\}$:
    \begin{equation}
        f\{Y(x)=y\}=\int_u\tilde{f}(y|u,x)\tilde{f}(u)du.
    \end{equation}
\end{enumerate}

\subsubsection{Null Treatment Method.} This method also depends on the \textit{equivalence assumption} and \textit{completeness assumption} mentioned above. Another key assumption called \textit{Null treatment}~\cite{miao2022identifying} is proposed as well. The cardinality of the intersection $\mathcal{C}\cap\mathcal{A}$ does not exceed $(|C|-q)/2$, where $|\mathcal{C}|$ is the cardinality of $\mathcal{C}$ and must be larger than the dimension of $U$.

Implementations of the causal inference is given below.
\begin{enumerate}
    \item Obtaining an arbitrary admissible joint distribution $\tilde{f}(x,u)$.
    \item Using the estimate $\tilde{f}(u|x)$ from Step (1), along with an estimate of $f(y|x)$, to solve the following equation for $\tilde{f}(y|u,x)$:
    \begin{equation}
        f(y|x)=\int_u\tilde{f}(y|u,x)\tilde{f}(u|x)du.
    \end{equation}

    \item Plugging the estimate of $\tilde{f}(u)$ from Step (1) and $\tilde{f}(y|u,x)$ from Step (2) into the equation below to estimate $f\{Y(x)\}$:
    \begin{equation}
        f\{Y(x)=y\}=\int_u\tilde{f}(y|u,x)\tilde{f}(u)du.
    \end{equation}
\end{enumerate}

\subsubsection{Likelihood Maximization}
Researchers also make attempt to estimate the effect of bundle treatment in the presence of hidden confounders under the framework of Structural Causal Model (SCM). They have proved that it is impossible to identify the joint effects of multiple treatments that is simultaneously taken if there is no restriction on the structure function~\cite{nandy2017estimating}. However, such influence could be estimated by introducing reasonable weak assumptions, such as the additive noise model. A simple parameter estimation method is proposed, where all the data from different regimes are pooled together in order to jointly maximize the combined likelihood.

A complementary question is also studied that how to estimate the causal effect of a single treatment while multiple treatments are adopted at the same time~\cite{jeunen2022disentangling}. Formally, given the samples which can deduce $\mathbb{E}[Y|T_i=t_i,T_j=t_j,X=x]$ and $ \mathbb{E}[Y|do(T_i=t_i,T_j=t_j),X=x]$, the purpose is to find how to learn the conditional average treatment effect $\mathbb{E}[Y|do(T_i=t_i),T_j=t_j,X=x]$ or $\mathbb{E}[Y|T_i=t_i,do(T_j=t_j),X=x]$. Researchers prove that this is not generally possible as well, unless there are non-linear continuous structural causal models with additive, multivariate Gaussian noise. They extend the Expectation Maximisation style iterative algorithm~\cite{nandy2017estimating} to disentangle the effects of each single treatment. Suppose the intervened treatments are $T_{int}\subseteq T$ and $T_{obs}\equiv T-T_{int}$, and then a causal query with $T_{int}$ could be decomposed as:
\begin{equation}
    \mathbb{E}[Y|C;do(X_{int});X_{obs}]=f_Y(C;X)+\mathbb{E}[U_Y|X_{obs}].
\end{equation}

\subsection{Conclusion and Discussion}
Under \textit{Unconfoundedness Assumption}, the key challenges of bundle treatment effect estimation are three-folds: (1) The complex confounding bias for exponential-level treatment assignments. (2) The need for a general hypothesis model for counterfactual prediction of all treatment groups. (3) The additional influence caused by the interactions of multiple treatments that are simultaneously taken. Although methods introduced above may only address some of these three challenges rather than all, the approaches they applied for specific problems still provide valuable insights. When it comes to the unobserved confounders, some methods (DIRECT and Deconfounder) try to tackle with a more difficult task that recovering the proxies only from the treatment assignments $T$ without any information of $X$. These approaches introduced in Section~\ref{sec:bundle-unobserve} mainly focus on the identifiablity of the causal effects, which are different from the research priorities of the methods in Section~\ref{sec:bundle-basic}.

\begin{table}[b]
\centering
\small
\caption{Datasets for continuous treatment.}
\vspace{-3mm}
\label{tab:data-continuous}
\begin{tabular}{ccc}
\toprule
\textbf{Dataset} & \textbf{Description} & \textbf{Link}  \\ \hline
TCGA & gene expression & \url{https://github.com/d909b/drnet} \\
MIMIC III  & ICU  & \url{https://mimic.mit.edu} \\
IHDP & infant health & \url{https://www.fredjo.com} \\
Medicare & PM $2.5$  & \url{https://doi.org/10.23719/1506014} \\
\bottomrule
\end{tabular}
\end{table}

\section{Datasets and Codes}~\label{sec:experiment}
In this section, we summarize the available datasets for multi-valued, continuous, and bundle treatments in Section~\ref{sec:dataset}. Methods whose code is open-sourced are concluded in Section~\ref{sec:code} as well. \textcolor{black}{Moreover, we develop a toolkit\footnote{\url{https://github.com/causal-machine-learning-lab/mlbt}} of causal inference for complex treatments}.

\subsection{Available Datasets}\label{sec:dataset}

Datasets applicable for evaluating methods of multi-valued treatment include Twins and News.

\textbf{Twins.} This dataset is collected from all births\footnote{\url{https://www.nber.org/research/data/linked-birthinfant-death-cohort-data}} in the USA between $1989$-$1991$, and only the twins weighing less than $2$kg are recorded without missing features~\cite{data-twins}. The outcome refers to the mortality after one year. It is originally utilized by models focused on binary treatment. After preprocessing, there are $11,400$ pairs of twins, along with $30$ covariates related to the parents, pregnancy and birth. MetaITE~\cite{meta} is the first one to extend twins dataset to the multi-valued problem, where $4$ treatments are considered: $T=0$ means lower weight and female sex; $T=1$ means lower weight and male sex; $T=2$ means higher weight and female sex; $T=3$ means higher weight and male sex.

\textbf{News.} It simulates the opinions of a media consumer exposed to multiple news items, which is generated from the NY Times corpus\footnote{\url{http://archive.ics.uci.edu/ml/datasets/bag+of+words}}. The purpose is to infer the individual treatment effects of obtaining more content from some specific devices on the reader’s opinions. In particular, each sample $x_i$ refers to news items represented by word counts, and outcome $y_i\in\mathbb{R}$ represents the reader’s opinions of the news. As for the intervention $t_i$, DRNet~\cite{DRNet} and MetaITE~\cite{meta} construct multiple treatments which correspond to various devices used to view the news items, including desktop, smartphone, newspaper, and tablet.

As concluded in Table~\ref{tab:data-continuous}, there are $4$ more datasets used in the case of continuous treatment.

\begin{table}[t]
\centering
\caption{Datasets for causal inference of bundle treatment.}
\vspace{-3mm}
\label{tab:data-bundle}
\resizebox{1\textwidth}{!}{
    \begin{tabular}{ccc}
    \toprule
    \textbf{Dataset} & \textbf{Description} & \textbf{Link}  \\ \hline
    NMES & smoking habits and medical expenses & \url{https://meps.ahrq.gov//mepsweb/data_stats/download_data_files.jsp} \\
    Movies  & actors and movie earnings & \url{https://www.kaggle.com/datasets/tmdb/tmdb-movie-metadata} \\
    Amazon & reviews and future sales & \url{https://cseweb.ucsd.edu/\~jmcauley/datasets.html\#amazon_reviews} \\
    CRISPR KO & causal genetic interaction & \url{https://ndownloader.figshare.com/files/ 25494359} \\
    \bottomrule
    \end{tabular}  
}
\end{table}

\textbf{Cancer Genomic Atlas (TCGA).} It contains $9,659$ observations with $20,531$ features from various types of cancers. DRNet~\cite{DRNet} makes use of this dataset for evaluation, where $3$ clinical treatments are taken into account including medication, chemotherapy, and surgery. The potential outcome studied here is the risk of cancer recurrence after receiving either of the treatment options.

\textbf{Medical Information Mart for Intensive Care (MIMIC) III~\cite{johnson2016mimic}}. This is a large, publicly available database, comprising information of $8,040$ patients who are admitted to critical care units at a large tertiary care hospital. Beside the $49$ features of a patient, it also includes a wide range of clinical data, including demographic information, vital signs, laboratory test results, medications, and clinical notes. DRNet~\cite{DRNet} and SciGAN~\cite{SciGAN} utilize MIMIC III dataset to study the causal effect of three antibiotics treatments on the arterial blood gas readings of the ratio of arterial oxygen partial pressure to fractional inspired oxygen.

\textbf{Infant Health and Development Program (IHDP)~\cite{IDHP}}. It is a longitudinal study that was conducted in the United States from $1985$ to $1993$. It contains data from $747$ infants with $25$ covariates. The infants were randomly assigned to either a treated group with high-quality educational and developmental services, or a control group with standard care. In BART~\cite{PEHE} and VCNet~\cite{VCNet}, this dataset is utilized to evaluate how much does the preschool education affects the IQ tests.

\textbf{Medicare~\cite{medicare}}. This is a collection of data on socioeconomic status for $2,132$ US counties, together with average annual cardiovascular mortality rate (CMR) and total PM $2.5$ concentration. This study spans over $21$ years ($1990$-$2010$) and covers more than $68.5$ million samples. Several works~\cite{wu2022matching,ren2021bayesian} focus on this dataset to estimate the long-term causal effect of PM $2.5$ on all-cause mortality under $18$ covariates. Note that the treatment ranges from $0.01$ to $30.92$, and $99\%$ of the data lies within the interval $(2.76, 17.16)$.

We summarize the available real-world or semi-synthetic datasets for bundle treatment in Table~\ref{tab:data-bundle}.

\textbf{National Medical Expenditures Survey (NMES).} It is a collection of data about smoking habits and medical expenses in a representative sample of the U.S. population. The dataset contains $9,708$ people and $8$ variables about each. In the implementation of Deconfounder (2018)~\cite{deconfounder-2018}, they only focus on the current marital status $a_{mar}$, the cumulative exposure to smoking $a_{exp}$, and the last age of smoking $a_{age}$.

\textbf{TMDB 5000 Movie Dataset.} It is a collection in Kaggle that contains $901$ actors (who appeared in at least $5$ movies) and the revenue for the $2,828$ movies they appeared in. The movies span $18$ genres and 58 languages. The purpose of Deconfounder (2018)~\cite{deconfounder-2018} applying this dataset is to study how much does an actor boost (or hurt) a movie’s revenue.

\textbf{Amazon-3C} and \textbf{Amazon-6C.} They are two semi-synthetic datasets from the Amazon review dataset in DIRECT~\cite{DIRECT}. In each dataset, $3$/$6$ categories of items are selected. Afterwards, the top $1,000$ products with most reviews are collected as instances in each category. The goal is to investigate the effect of the keywords in reviews on the future sales of each product. Specifically, treatments here refer to $3$ key words derived from the reviews, and the potential outcome is the simulated future amount of sales of each product. Confounders are the latent attributes of the products, which are generated by training a neural network to fit the treatment assignment.

\textbf{CRISPR Thee-way Knockout (CRISPR KO).} It is a benchmark dataset consists of real-world experimental data collected in a systematic multi-gene knockout screen~\cite{zhou2020three}. As pointed out in NCoRE~\cite{NCoRE}, it is particularly challenging for causal inference on this dataset because of the complex underlying biological process, which leads to the large number of potential treatment combinations, high-dimensional covariate space, and the sparsity of labelled data available.

\subsection{Open-source}\label{sec:code}
The available codes for causal inference with complex treatments are summarized in Table~\ref{tab:codes}, including multi-valued, bundle, and continuous settings. Considering that the IV methods can be naturally developed to solve the causal estimation of continuous treatment, readers can also refer to the toolkit\footnote{\url{https://github.com/causal-machine-learning-lab/mliv}} of IVs methods that is reviewed in the survey of IV~\cite{iv-survey}.

\begin{table*}[h]
\centering
\caption{Available codes of methods for complex treatment.}
\vspace{-3mm}
\label{tab:codes}
\resizebox{1\textwidth}{!}{
    \begin{tabular}{cccc}
    \toprule
    \textbf{Category} & \textbf{Method} & \textbf{Language} & \textbf{Link}  \\ \hline
    \multirow{3}{*}{Multi-valued}
    & GANITE & python & \url{https://github.com/vanderschaarlab/mlforhealthlabpub/tree/main/alg/ganite} \\
    & \multirow{2}{*}{CTS} & python & \url{https://github.com/Ibotta/mr_uplift} \\
    & & R & \url{https://github.com/Matthias2193/APA} \\
    \hline\hline
    \multirow{3}{*}{Bundle} 
    & SCP & python & \url{https://github.com/ZhaozhiQIAN/Single-Cause-Perturbation-NeurIPS-2021} \\
    & Deconfounder (2018) & python & \url{https://github.com/blei-lab/deconfounder_tutorial} \\
    & Null Treatment & R & \url{https://github.com/JiajingZ/CopSens} \\
    \hline\hline
    \multirow{11}{*}{Continuous}
    & TR & python & \url{https://github.com/claudiashi57/dragonnet} \\
    & SciGAN & python & \url{https://github.com/ioanabica/SCIGAN} \\
    & DRNet & python & \url{https://github.com/d909b/drnet} \\
    & VCNet & python & \url{https://github.com/lushleaf/varying-coefficient-net-with-functional-tr} \\
    & KPV & python & \url{https://github.com/afsaneh-mastouri/kpv} \\
    & PMMR & python & \url{https://github.com/yuchen-zhu/kernel_proxies} \\
    & DFPV & python & \url{https://github.com/liyuan9988/deepfeatureproxyvariable} \\
    & VMM & python & \url{https://github.com/CausalML/VMM} \\
    & Kernel IV & matlab & \url{https://github.com/r4hu1-5in9h/KIV} \\
    & CB-IV & python & \url{https://github.com/anpwu/CB-IV} \\
    & DeepGMM & python & \url{https://github.com/CausalML/DeepGMM} \\
    & AGMM & python & \url{https://github.com/vsyrgkanis/adversarial_gmm} \\
    \bottomrule
    \end{tabular}
}
\vspace{-3mm}
\end{table*}

\section{Future Work and Discussion}~\label{sec:discussion}
In this section, we will point out some limitations of the existing methods, discuss the fundamental challenges for different settings, and give some ideas of the directions for future works.

\subsection{Multiple Discrete Treatments}
Multi-valued treatment and bundle treatment can be unified into multiple discrete treatments, and every single treatments are mutually exclusive with each other for the former while not for the latter. We conclude several challenges that deserve further exploration.

\subsubsection{How to control the confounders among different treatments?}
The key objective of causal inference is to eliminate the confounding bias, and this problem gets more challenging in the settings of multi-valued treatment and bundle treatment. Recently, researchers are bound up in leaning representations for treatment ($T$) and confounders ($X$ and $U$).

From the perspective of treatment representation, the mainstream includes invariant embedding and variational embedding by VAE. The former method applies limitations like MMD to control the distributions of different groups, while the latter does not address it explicitly. A further question is that if there is a need to consider the similarity among treatments. This concern expands to the following aspects:
\begin{enumerate}
    \item Some single treatments may be similar to each other.
    \item Bundle treatments that are highly overlapped may be mutually similar as well.
    \item How to trade-off the prominent different treatment elements in minority of two bundle treatments while the remaining parts are quite overlapped in majority.
\end{enumerate}

When it comes to the representation of confounders, existing methods include treatment assignment decomposition and $\beta$-VAE. Interpretability and disentangled capability of them are still in doubt. Besides, there are more concerns need further discussion:
\begin{enumerate}
    \item Different treatments correspond to different confounders.
    \item Similar treatments share the same confounders.
    \item Whether the embedding of a bundle treatment simply equals to the union of all the involved single treatments.
\end{enumerate}

Even worse, there exists a fatal problem in models inspired by advanced AI techniques like VAE or DA, i.e. the proof in many of them is proposed for a lower bound rather than an unbiased estimate for ATE or ITE. Strictly speaking, what they pursue in the present stage is not guaranteed to be the exact causal effect.

\subsubsection{How to measure the interactions among multiple treatments?}
This is a unique challenge for bundle treatment setting, where multiple treatments could be received in the meantime. The factual outcome in observational data corresponds to the assigned bundle treatment, then how to exfoliate the actual effect of every single treatment? Moreover, it is open to discuss either using a general model for counterfactual prediction or splitting the entire outcome as combinations of single treatment effects together with a correction term. This concern makes sense in many real-world scenarios. For instance, $Y(antillergic\ pill) + Y(immunity\ injection) > sum$, $Y(flu\ medicine)+Y(mist\ spray)=sum$, and $Y(ibuprofen)+Y(compound\ paracetampol)<sum$.

\subsubsection{How to tackle with data sparsity?}
When the number of optional treatments gets larger, the entire treatment space also grows exponentially. Therefore, data sparsity will become challenging in real-world applications. 
SCP~\cite{SCP} that designed in a manner of data augmentation is a feasible solution. Besides, MetaITE~\cite{meta} also works by modeling the group with few samples as a targeted domain. Methods from other communities like DA and meta learning may give some new insights into causal inference with large treatment space.

\subsection{Continuous Treatment}
The development of causal inference with continuous treatment is much better than that of multi-valued and bundle treatment. This gap is partly due to the solid foundation of IV in binary treatment, which can be easily extended as a solution of the continuous treatment problem. 

However, current literature with respect to continuous treatment mainly focus on the potential outcome framework, which heavily rely on the discovering the causal graph structure, namely causal discovery. It is challenging to ensure the identifiability of causal effects in a complex graph structure, and it is tough to specify the causal graph from observational data.

\section{Conclusion}~\label{sec:conclusion}
In this survey, we provide a comprehensive review of the existing methods of causal inference with complex treatment settings, including multi-valued, continuous, and bundle treatment. We clarify the problem setting, common notifications, and basic assumptions in Section~\ref{sec:preliminary}. Methods tackling with binary treatment are also introduced as background knowledge in Section~\ref{sec:binary}. We discuss the methods focused on multi-valued treatment in Section~\ref{sec:multi}, those following the unconfoundedness assumption are organized in the first part while those considering unobserved confounders are described in the second part. It is the same for methods with continuous treatment in Section~\ref{sec:continuous} and methods with bundle treatment in Section~\ref{sec:bundle}. We then comb through the available datasets and open-sourced codes from cluttered literature in Section~\ref{sec:experiment}. To the best of our knowledge, it is the first work that summarizes all the aforementioned information and unify the three kinds of treatments into \textit{complex treatments}. In Section~\ref{sec:discussion}, we discuss the challenges encountered in these new settings and some potential directions for future explorations.


\bibliographystyle{ACM-Reference-Format}
\bibliography{sample-base}


\end{document}